\documentclass[onecolumn,sort&compress,numbers]{els-mrw} % For numbered references Style 

\usepackage{amsmath,amssymb,amsfonts,amsthm,makeidx,graphicx}
\usepackage{txfonts}
\usepackage{helvet}
\usepackage{float}
\usepackage{tcolorbox}
\usepackage{tikz}
\usepackage[compat=1.1.0]{tikz-feynman}
\usepackage[colorlinks=true, linkcolor=blue, citecolor=blue, urlcolor=blue]{hyperref}

\newcommand{\mdp}{m_{A'}}

\begin{document}

%%%%%%%%%%%%%%%%%%%%%%%%%%%%%%%%%%%%%%%%%%%%%%%%%%%%%%%%%%%%%%%%
%% the following items are mandatory: 
%% - title
%% - author names
%% - affiliation details
%% - abstract
%% - keywords

%% Precise, concise, and informative description of the focus of this work. Avoid abbreviations and formulae in the title
%\chapter{Article title (template for all chapters in Sections 1-5: General Concepts, Hadron Physics, EW Physics, Neutrino Physics, BSM)}\label{chap1}
\chapter{The Dark Photon: a 2026 Perspective}\label{chap1}

%% All author names and affiliations, and email address for corresponding author
\author[1,2,3]{Andrea Caputo}%
\author[4]{Rouven Essig}%

\address[1]{\orgname{CERN}, \orgdiv{Department of Theoretical Physics}, \orgaddress{Esplanade des Particules 1, P.O. Box 1211, Geneva 23, Switzerland}}
\address[2]{\orgname{``Sapienza'' Universita` di Roma \& Sezione INFN}, \orgdiv{Dipartimento di Fisica}, \orgaddress{Piazzale Aldo Moro 5, 00185, Roma, Italy}}
\address[3]{\orgname{Department of Particle Physics and Astrophysics}, \orgaddress{Weizmann Institute of Science, Rehovot 7610001, Israel}}
\address[4]{\orgname{Stony Brook University}, \orgdiv{C.N.~Yang Institute for Theoretical Physics}, \orgaddress{Stony Brook, NY 11794, USA}}
%\address[1]{\orgname{Name of Institution}, \orgdiv{Division or Department}, \orgaddress{Address of Institution}}

\articletag{Chapter Article tagline: update of previous edition, reprint.}

\maketitle

%%%%%%%%%%%%%%%%%%%%%%%%%%%%%%%%%%%%%%%%%%%%%%%%%%%%%%%%%%%%%%%%
%% the following item is optional: 
%% - System of abbreviations/terms/symbols used in the specific field of study/community. List and define
\begin{glossary}[Nomenclature]
	\begin{tabular}{@{}lp{34pc}@{}}
        Dark photon & also called hidden photon, heavy photon, massive photon, $U$-boson\\
        SM & Standard Model\\
        BSM & Beyond the Standard Model\\
        CMB & Cosmic microwave background\\
        BBN & Big Bang Nucleosynthesis
	\end{tabular}
\end{glossary}

%%%%%%%%%%%%%%%%%%%%%%%%%%%%%%%%%%%%%%%%%%%%%%%%%%%%%%%%%%%%%%%%
%% the following item is mandatory: 
%% List of the key points and topics a reader can expect to learn from this chapter 
\section*{Objectives}
\begin{itemize}
	\item Dark photons provide an important portal to a dark sector.  They are ubiquitous in extensions of the Standard Model, and they provide a simple possibility for how dark matter particles interact with ordinary matter and/or with themselves.  
    \item The phenomenology of dark photons depends sensitively on their mass.  Dark photons with a mass above $\sim$1~MeV can decay quickly to electron-positron pairs and (if their mass is sufficiently large) to other Standard Model particles, while dark photons with a mass below $\sim$1~MeV are naturally long-lived.  We therefore divide up this review into these two mass ranges. 
    \item There are numerous probes of dark photons.  This includes searches using accelerators, such as collider, fixed-target, and beam-dump experiments; astrophysical probes, such as supernova; and cosmological probes, such as Cosmic Microwave Background spectral distortions and Big Bang Nucleosynthesis. 
\end{itemize}

%%%%%%%%%%%%%%%%%%%%%%%%%%%%%%%%%%%%%%%%%%%%%%%%%%%%%%%%%%%%%%%%
%% the following item is mandatory: 
%% 100-150 word summary of the chapter
\begin{abstract}[Abstract]
	We give a pedagogical overview of dark photons.  We describe the theory and their importance in particle physics research, and discuss searches using laboratory, astrophysical, and cosmological probes. 
    %Text of your abstract e.g.: We give a pedagogical introduction...(100-150 words)
\end{abstract}

%% 5-10 words that embody the key topics in the chapter. What terms would someone put into a search engine if they were looking for a chapter like this?
\begin{keywords}
 	%please enter 5 keywords as follows:
 	dark photon\sep dark sector\sep hidden photon \sep portals \sep dark matter
\end{keywords}

%%%%%%%%%%%%%%%%%%%%%%%%%%%%%%%%%%%%%%%%%%%%%%%%%%%%%%%%%%%%%%%%
\section{Introduction}

A major goal of particle physics is to identify the fundamental constituents of matter and how these interact. This endeavor has resulted in the Standard Model (SM) of particle physics, which describes much of the known physical world.  However, the SM is also known to be an incomplete theory, as it does not describe the particles that make up the dark matter in the Universe, which is known to comprise $\sim$85\% of the matter density.  The SM also does not explain, for example, the origin of the neutrino masses, the origin of the baryon asymmetry, the masses and interactions of the various flavors of quarks, and why the Higgs mass is so much less than the Planck mass.  

A pertinent question is then what particles might exist Beyond the Standard Model (BSM) and, in particular, which ones might be connected to the dark matter. There are many possibilities, but among the simplest and richest possibilities is a dark photon. The dark photon mediates a new force.  In this review, we will focus on massive dark photons, and we will consider dark photons that interact with electromagnetically charged particles through ``kinetic mixing'' (to be described below).  For a range of masses and couplings, such dark photons can constitute the dark matter, but they may also just constitute a BSM particle without being part of the dark matter.  We will also discuss the possibility that dark photons are a mediator of a force among dark matter particles.  

Dark photons are an important topic in particle physics and appear in many hundreds of papers. The reasons for their popularity are multifold. First, since they are the vector boson that is part of an U(1) gauge group, they are ubiquitous in BSM theories that attempt to extend the SM, since such extensions often contain additional U(1) gauge groups.  Second, as we will discuss in \S\ref{sec:theory}, they can play an important role in physics at low energies, even if their coupling to the SM is generated at very high energies. Third, they have a rich phenomenology, as they can be probed in numerous ways, and this richness has led to many studies.   Fourth, new experiments and data are needed to probe parts of the viable parameter space, which has led to much activity and exciting interactions between theorists and experimentalist~\cite{Essig:2022yzw}.  Fifth, dark photons provide an exciting ``portal'' to a dark sector, and, in particular, dark matter may be connected to ordinary matter through the dark photon.  Sixth, they are theoretically quite simple, and often provide a useful approximation to more general dark-sector theories. 

In our overview, we will not provide a detailed description of all phenomena that dark photons can impact, nor will we provide an exhaustive reference list. Some reviews in which dark photons and dark sectors have been discussed include~\cite{Jaeckel:2010ni,Essig:2013lka,Alexander:2016aln,Battaglieri:2017aum}.  The importance of dark photons, their possible connection to dark matter, and dark sectors more generally have been recognized in numerous papers, including~\cite{Boehm:2003hm,Strassler:2006im,Pospelov:2007mp,Arkani-Hamed:2008hhe,Pospelov:2008jd}.

%%%%%%%%%%%%%%%%%%%%%%%%%%%%%%%%%%%%%%%%%%%%%%%%%%%%%%%%%%%%%%%%
%% the following item is optonal: 
%% - Single figure visually illustrating the key topic/method/outcome described in the chapter
%\begin{figure}[h]
%	\centering
%	\includegraphics[width=7cm,height=4cm]{blankfig}
%	\caption{Optional: Single figure visually illustrating the key topic/method/outcome described in the chapter. 
%		     Please add here some text explaining the pic...}
%	\label{fig:titlepage}
%\end{figure}

\section{Dark photon theory} \label{sec:theory}

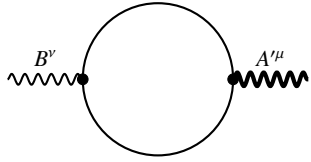
\begin{figure}[t]
\centering
\begin{tikzpicture}
  % Left external wavy line (thinner)
  \draw[decorate, decoration={snake, amplitude=2pt, segment length=5pt}, thick] 
        (-2,0) -- (-1,0) node[midway, yshift=8pt] {\(B^\nu\)};
        
  % Right external wavy line (thicker)
  \draw[decorate, decoration={snake, amplitude=2.5pt, segment length=5pt}, ultra thick] 
        (1,0) -- (2,0) node[midway, yshift=8pt] {\(A'^\mu\)};
  
  % Fermion loop
  \draw[thick] (-1,0) arc[start angle=180,end angle=0,radius=1cm];
  \draw[thick] (1,0) arc[start angle=0,end angle=-180,radius=1cm];

  % Interaction vertices
  \filldraw[black] (-1,0) circle (2pt);
  \filldraw[black] (1,0) circle (2pt);
\end{tikzpicture}
\caption{Loop diagram generating kinetic mixing between $B^\nu$ and $A'^\mu$ via a charged fermion loop. In this case, with only one loop generating the mixing operator, one expects $\epsilon \sim 10^{-2}-10^{-1}$, much above the present experimental constraints, as we will see. Thus, one typically needs the suppression from multi-loop processes. Kinetic mixing through gravity alone, for example, requires at least six loops and leads to $\varepsilon \sim 10^{-13}$~\cite{Gherghetta:2019coi}.}
\label{Fig:loop_kinetic_mixing}
\end{figure}

We begin by reviewing how a dark photon, which we denote as $A'$, can interact with ordinary matter via kinetic mixing.  In addition to the SM particle content, we consider an abelian gauge theory $U(1)_D$, whose mediator is the dark photon field $A'_\mu$.  The relevant terms in the Lagrangian are 
\begin{equation}\label{eq:A'-Lagrangian}
\mathcal{L} \supset -\frac{1}{4} F^{Y,\mu\nu}F^Y_{\mu\nu} - \frac{1}{4} W_3^{\mu\nu}W_{3,\mu\nu} -
\frac{1}{4} F'^{,\mu\nu}F'_{\mu\nu} + \frac{\epsilon}{2\cos\theta_W} F^{Y,\mu\nu} F'_{\mu\nu} + m_{A'}^2 A'^{\mu} A'_{\mu}\ , 
\end{equation}
where $F^Y_{\mu\nu}=\partial_{\mu}B_{\nu}-\partial_{\nu}B_{\mu}$ is the field strength tensor of the SM hypercharge vector boson $B_\mu$ of the $U(1)_Y$ gauge theory, $W_{3,\mu\nu}=\partial_{\mu}W_{3,\nu}-\partial_{\nu}W_{3,\mu}$ is the field strength tensor of the SM SU(2)$_L$ gauge field $W_3$ that together with $B_\mu$ forms the photon and $Z$ boson mass eigenstates, $F'_{\mu\nu}=\partial_{\mu}A'_{\nu}-\partial_{\nu}A'_{\mu}$ is the field strength tensor of the $U(1)_D$ gauge theory, $\mdp$ is the mass of the dark photon, and $\theta_W$ is the Weak mixing angle.  The term $\frac{\epsilon}{2\cos\theta_W}  F^{Y,\mu\nu} F'_{\mu\nu}$ is the kinetic mixing term~\cite{Holdom:1985ag,Galison:1983pa,Dienes:1996zr} and induces a mixing between the kinetic energy terms of the SM hypercharge and dark photon field.  This operator can be generated by loops of new massive particles that couple to both the dark photon and hypercharge gauge bosons, producing a range of plausible values for the kinetic-mixing parameter $\epsilon$ of $\sim 10^{-8}-10^{-2}$ (see, e.g., Fig.~\ref{Fig:loop_kinetic_mixing}), although smaller values are also possible in string theory constructions~\cite{Abel:2008ai}. It is important to note that the kinetic-mixing term is a dimension four operator, and hence it is not suppressed by the mass scale at which it is generated (in other words, even if this operator is generated by particles with a mass at the scale of grand unification ($\sim 10^{16}$~GeV), $\epsilon$ is not suppressed by that scale and can still be sizeable.  This makes the dark photon model particularly interesting, as it could provide the dominant connection to a dark sector.

The last term in Eq.~(\ref{eq:A'-Lagrangian}) represents the mass of the dark photon.  The mass can be generated by a (dark) Higgs mechanism, in which case the dark sector would contain also a dark Higgs boson. Alternatively, the mass can be generated by the St\"uckelberg mechanism~\cite{Feldman:2007wj,Feng:2014eja,Feng:2014cla}. The mass can take on a wide range of values, and in this review, we will consider values of $m_{A'}$ in the range $\sim10^{-18}$~eV to 100~GeV. We note that we mostly ignore mass mixing between the dark Higgs and Standard Model Higgs in this review. However, this topic has been discussed in several papers, including e.g.~\cite{Curtin:2014cca}.

\begin{figure}[b]
	\centering
\includegraphics[width=.49\linewidth]{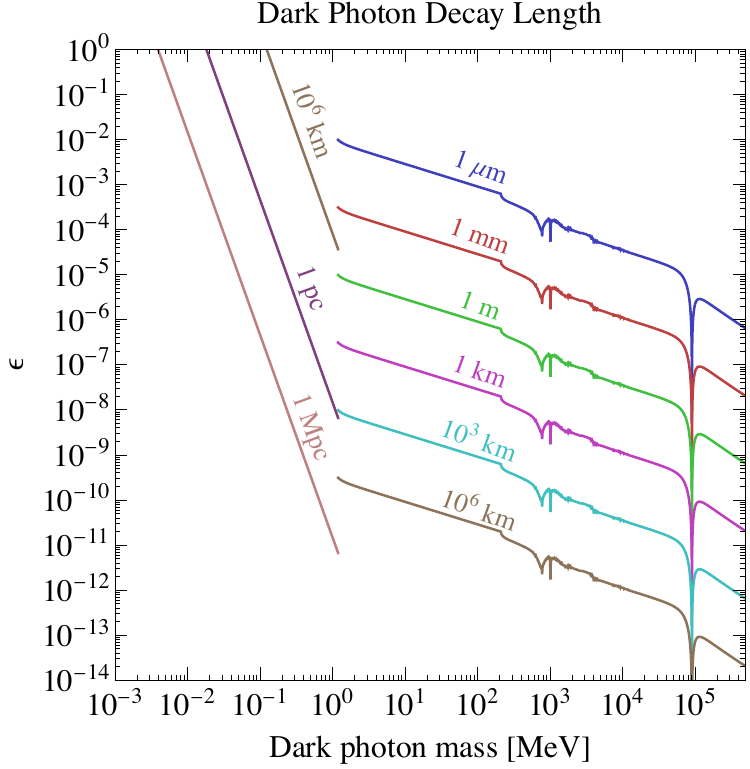}
~~
\includegraphics[width=.49\linewidth]{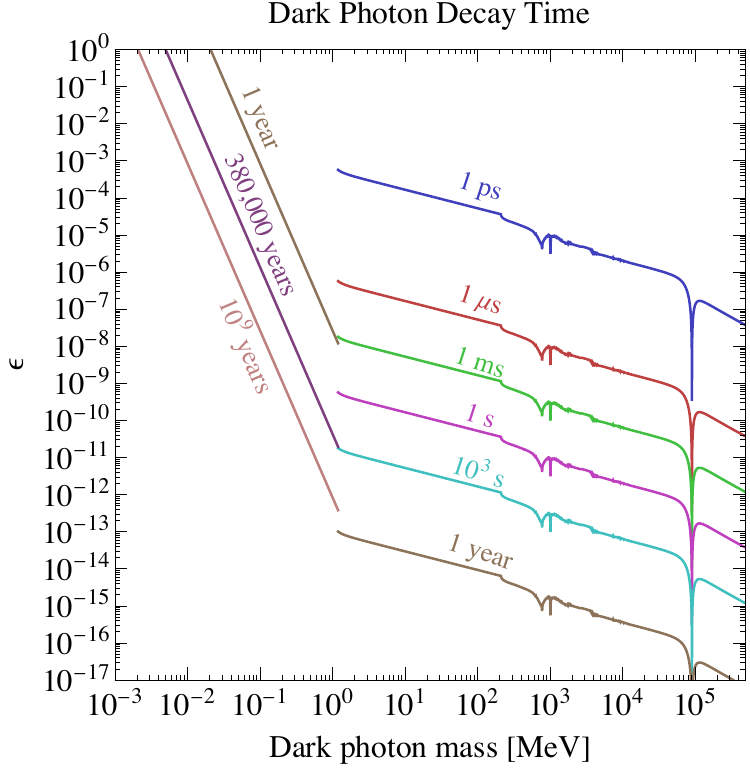}\\
\caption{Dark photon decay length (left) and decay time (right).  These decay lengths and times are in the rest frame of the particle; a boost factor of $E_{A'}/\mdp$, where $E_{A'}$ is the energy of the dark photon, needs to be added to obtain the values in the laboratory frame. We note the sharp transition around $\mdp\sim 2m_e\sim$1~MeV. 
}
	\label{fig:DecayLength-Lifetime}
\end{figure}

The Lagrangian in Eq.~(\ref{eq:A'-Lagrangian}) is in the interaction basis, and it is useful to re-write it in the mass-eigenstate basis.  For this, we follow the procedure discussed in e.g.~\cite{Baumgart:2009tn}, and refer the reader to, e.g., \cite{Curtin:2014cca} for a more detailed discussion. As in the SM, we define the $Z$-boson ($Z_\mu$) and SM photon ($A_\mu$) mass eigenstates, $Z_\mu = \cos\theta_W W_{3,\mu} - \sin\theta_W B_\mu$ and $A_\mu = \sin\theta_W W_{3,\mu} - \cos\theta_W B_\mu$. Eq.~(\ref{eq:A'-Lagrangian}) can then be written as 
\begin{equation}\label{eq:A'-Lagrangian-mass-basis}
\mathcal{L} \supset -\frac{1}{4} F^{\mu\nu}F_{\mu\nu} - \frac{1}{4} Z^{\mu\nu}Z_{\mu\nu} -
\frac{1}{4} F'^{,\mu\nu}F'_{\mu\nu} + \frac{\epsilon}{2\cos\theta_W} (\cos\theta_W F_{\mu\nu} - \sin\theta_W Z_{\mu\nu}) F'_{\mu\nu} + m_{A'}^2 A'^{\mu} A'_{\mu}\ , 
\end{equation}
where $F_{\mu\nu}$ and $Z_{\mu\nu}$ are the field strengths for the photon and $Z$-boson, respectively. 
We now redefine the fields of the photon and dark photon,
\begin{eqnarray}\label{eq:field-redefinition}
    A_\mu &\to& A_\mu - \epsilon A'_\mu \\
    A'_\mu &\to& A'_\mu + \epsilon \tan\theta_W Z_\mu\,,
\end{eqnarray}
where we denote the original and modified fields by the same symbol to avoid requiring new notation. The kinetic mixing between the photon and the dark photon then vanishes, as does the kinetic mixing term between the dark photon and the $Z$-boson (up to $\mathcal{O}(\epsilon^3)$).  From Eq.~(\ref{eq:A'-Lagrangian-mass-basis}) it seems that no interaction remains between the dark photon and SM particles.  However, this is not so, since the field redefinition in Eq.~({\ref{eq:field-redefinition}}) also affects the SM currents (i.e., the interactions between the SM gauge bosons and fermions).  In particular, the electromagnetic and $Z$-boson currents $\mathcal{L}_{{\rm EM},Z} =  - e A_\mu J_{\rm EM}^\mu - g Z_\mu J_{\rm Z}^\mu$ are modified to 
\begin{equation}\label{eq:A'-coupling-EM-current}
\mathcal{L}_{{\rm EM},Z} + \epsilon e A'_\mu J_{\rm EM}^\mu + \mathcal{O}\left((m_{A'}/m_Z)^2\right) \epsilon A' J_{\rm Z}^\mu\ , 
\end{equation}
generating couplings between the dark photon and these two currents. 
We will almost exclusively consider dark photon masses $m_{A'}\ll m_Z$, in which case the interaction between the dark photon and $Z$-current is suppressed.  The dominant coupling that remains is between the dark photon and the electromagnetic current, which is suppressed by $\epsilon$. 

Eq.~(\ref{eq:A'-coupling-EM-current}) is central to understanding the phenomenology of dark photons.  Since they couple to electrically charged particles, they can be produced by interactions among such particles and (depending on the dark photon mass) decay to them.  
The lightest electrically-charged particles in the SM are the electron and positron (with a mass of $m_e\simeq 0.511$~MeV), and hence searches for dark photons are drastically different for $\mdp>2m_e\sim$1~MeV compared to $\mdp<2m_e$.  In particular, above 1~MeV, the dark photons can decay to (at least) electron-positron pairs, while below 1~MeV, the dominant decay is to three photons through a loop of charged particles.  The decay lifetime and decay length are just much larger for dark photon masses below 1~MeV compared to dark photon masses above 1~MeV, see Fig.~\ref{fig:DecayLength-Lifetime}. This is the reason we consider these two mass ranges separately in subsequent sections. 

\begin{figure}[t]
	\centering
	\includegraphics[width=12cm,height=9cm]{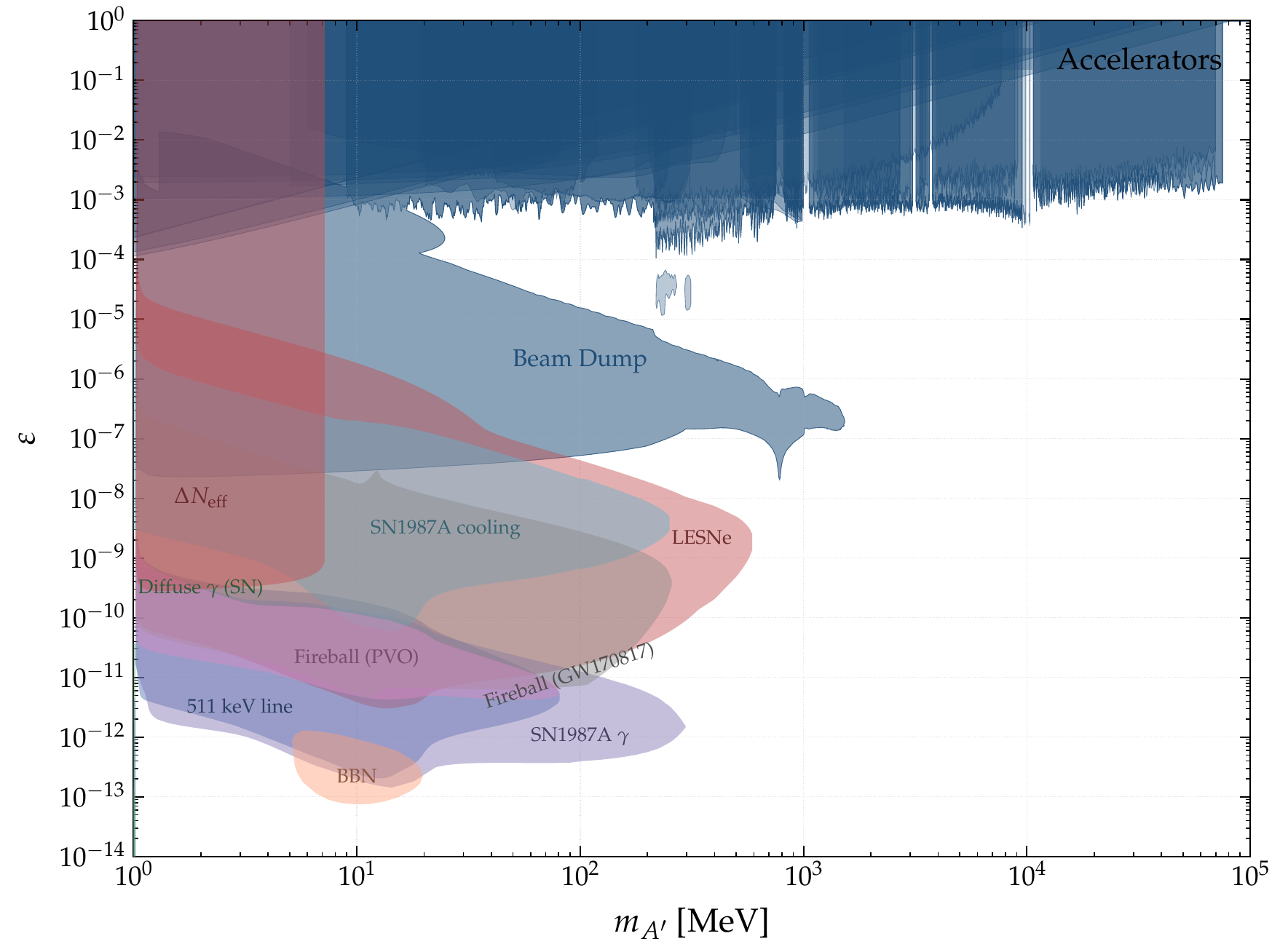}
\caption{Dark photon bounds for masses above the MeV, including constraints from colliders (dark blue and blue), supernovae and cosmology. Collider limits were recast using the \texttt{DarkCast} tool~\cite{DarkCast} and~\cite{Kyselov:2024dmi}. Astrophysical and cosmological bounds are based on results from Ref.~\cite{Caputo:2025avc}, where all constraints were computed from scratch.}
	\label{fig:BoundsAboveMeV}
\end{figure}

\section{Dark photons with mass above 1~MeV}

In this section, we discuss the searches and constraints on dark photons with masses above 1~MeV.  Fig.~\ref{fig:BoundsAboveMeV} summarizes the three types of searches that exist (accelerator-based probes, cosmology, and supernova) and the parameter regions they probe. 

\subsection{Precision measurements}

\begin{figure}[h]
\centering
\includegraphics[width=.35\linewidth]{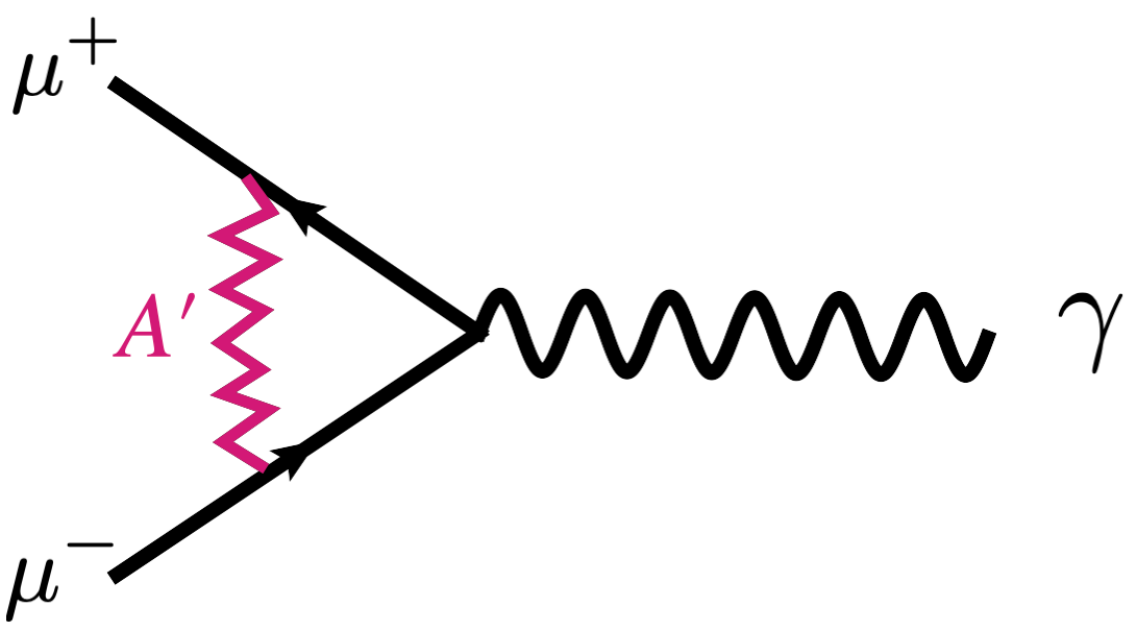}
\caption{The dark photon contributes to the anomalous magnetic moment of the muon through this loop diagram~\cite{Pospelov:2008zw}.}
\label{Fig:amu}
\end{figure}

Among the first constraints derived on dark photons made use of precision measurements of the anomalous magnetic moment of the electron and muon~\cite{Pospelov:2008zw}, since the dark photon impacts their values through the diagram shown in Fig.~\ref{Fig:amu}.  For example, for dark photon masses well below the muon mass, the contribution to anomalous magnetic moment of the muon is~$\sim\alpha\epsilon^2/2\pi$.  The discrepancy between the measured and expected SM value of the anomalous magnetic moment of the muon initially even provided a preferred region in the $\epsilon$ versus $\mdp$ parameter space for a few years, which has since then been ruled out by fixed-target and electron-positron collider data (see \S\ref{subsec:>1MeVaccelerators}).  

\subsection{Searches at accelerators}\label{subsec:>1MeVaccelerators}

For $\mdp>2m_e$, the dark photons can decay to electromagnetically charged particles.  A good approximation for the decay rate is given by~\cite{Bjorken:2009mm}
\begin{equation}
\Gamma(A' \to f^+ f^-) = N_{\rm eff}(\mdp) \ \frac{1}{3} \alpha_{\rm em} \, \varepsilon^2 m_{A'} \sqrt{1 - \frac{4m_f^2}{m_{A'}^2}} \left( 1 + \frac{2m_f^2}{m_{A'}^2} \right)\,. %\simeq 3.7 \times 10^{-3} \, \text{s}^{-1} \Big(\frac{\epsilon}{10^{-11}}\Big)^2 \Big(\frac{m_{A'}}{10 \, \text{MeV}}\Big),
\end{equation}
Here $N_{\rm eff}(\mdp)=1$ for $2m_e \le m_{A'} < 2m_\mu$, since only decays to $e^+e^-$ are possible.  For $m_{A'} \ge 2m_\mu$, we can use the measured ratio of $R\equiv\sigma(e^+e^-\to {\rm hadrons})/\sigma(e^+e^-\to \mu^+\mu^-)$~\cite{ParticleDataGroup:2024cfk}, so that $N_{\rm eff} (\mdp) = 2+R(\mdp)$. For a dark photon of energy $E_{A'}$, we can translate this to a decay length in the laboratory frame (we assume that $\mdp$ is not close to a threshold)
\begin{equation}
\ell_{\rm lab} = \frac{E_{A'}c}{m_{A'}\Gamma} \simeq 0.8~{\rm cm} \ \frac{1}{N_{\rm eff}} \frac{E_{A'}}{10~{\rm GeV}}\ \left(\frac{10^{-4}}{\epsilon}\right)^2\ \left(\frac{100~{\rm MeV}}{m_{A'}}\right)^2\,.
\end{equation}
Depending on the values of $\mdp$ and $\epsilon$, the decay can be prompt, produce a small displaced vertex, or be long-lived.  Different search strategies are thus needed to cover the entire $\epsilon$ versus $\mdp$ parameter space. 

\begin{figure}[b!]
\centering
\includegraphics[width=.49\linewidth]{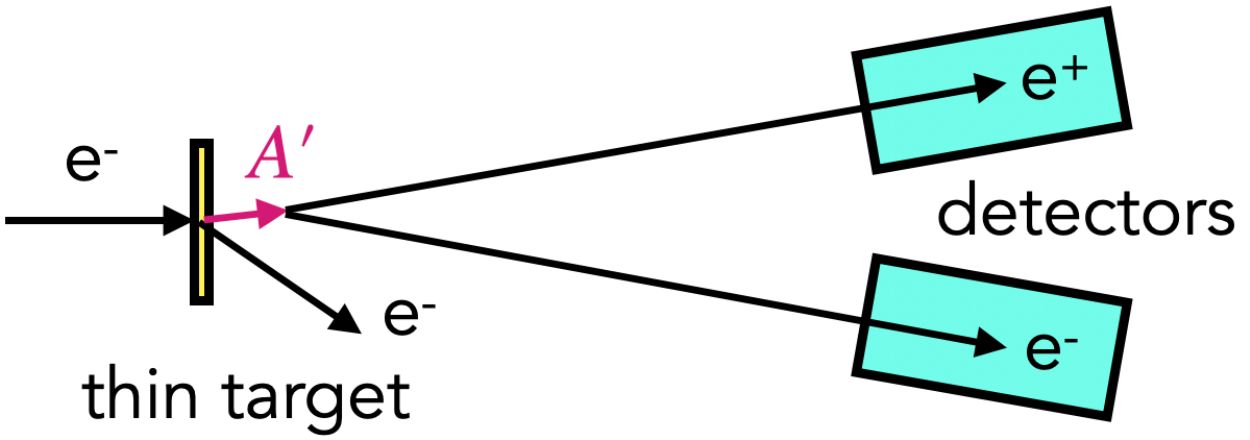}
~~
\includegraphics[width=.49\linewidth]{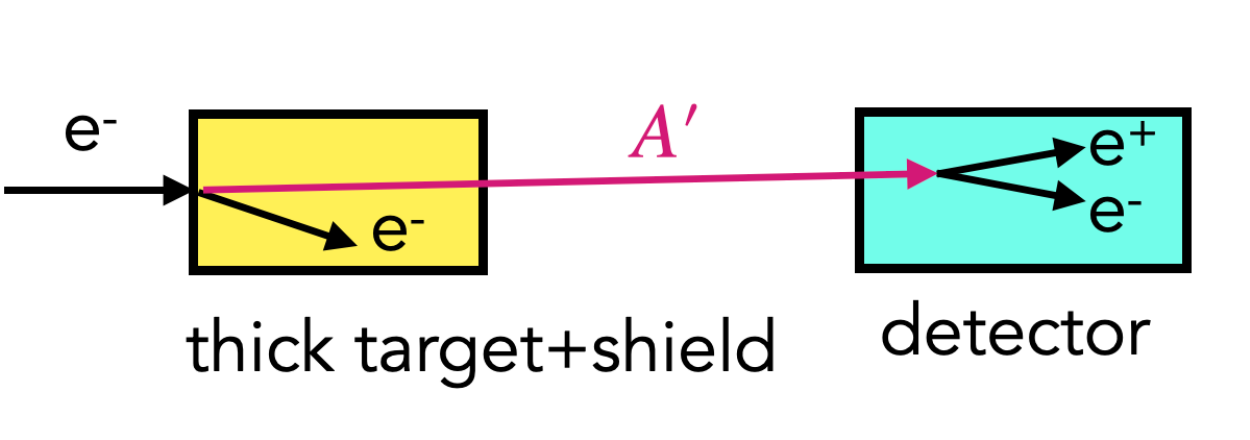}\\
\vskip 5mm
\includegraphics[width=.49\linewidth]{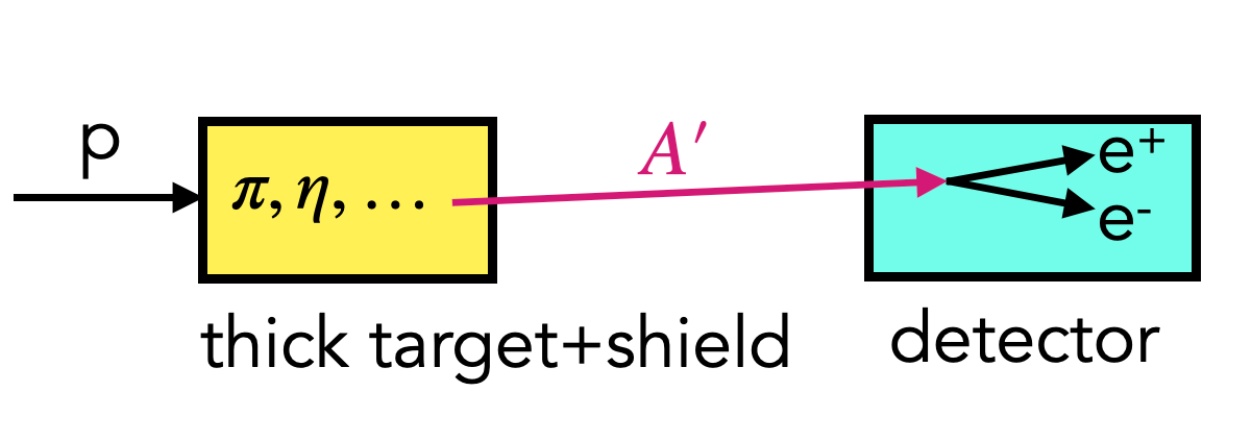}
\includegraphics[width=.49\linewidth]{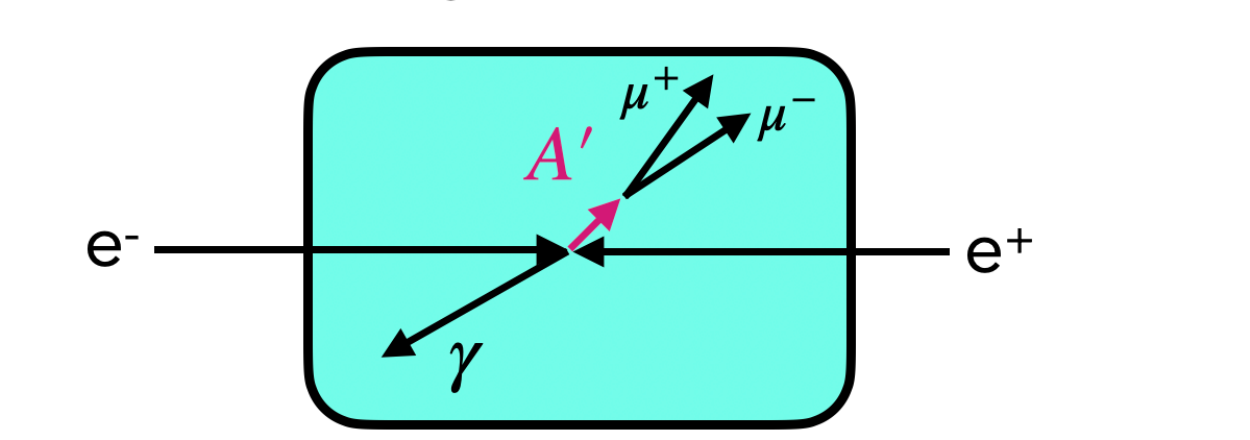}
 \caption{Dark photons can be produced at electron fixed-target experiments (top left), electron beam dump experiments (top right), proton beam-dump experiments (bottom left), and electron-positron colliders. They can subsequently decay to fermion-antifermion pairs.} 
 \label{fig:production-A'-schematic}
\end{figure}

\begin{figure}[t!]
\centering
\includegraphics[width=.35\linewidth]{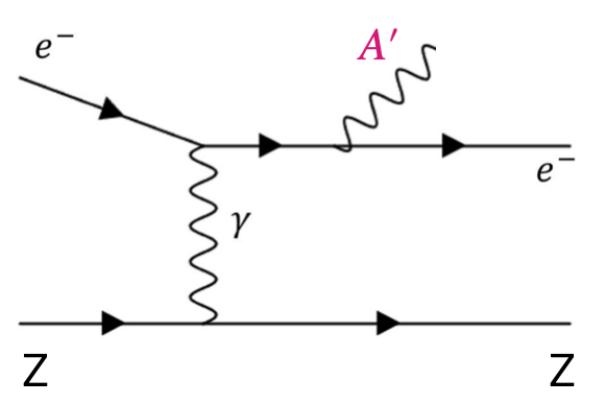}
\hskip 2cm
\includegraphics[width=.2\linewidth]{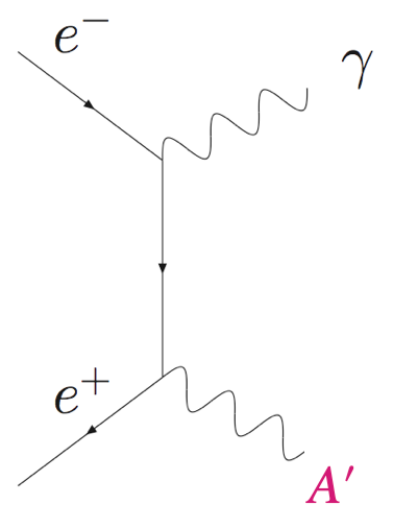}
 \caption{Examples of Feyman diagrams that contribute to dark photon production at electron-positron colliders (left) and at fixed-target experiments (right, where $Z$ indicates a nucleus of charge $Z$).  } 
 \label{fig:production-diagrams}
\end{figure}

\textbf{Electron-beam fixed-target experiments}, in which an electron beam impinges on a (thin) target, are a powerful probe of dark photons in the MeV to GeV mass range~\cite{Bjorken:2009mm,Reece:2009un,Freytsis:2009bh}, see Fig.~\ref{fig:production-A'-schematic} (left). The dark photon is produced through a diagram as shown in Fig.~\ref{fig:production-diagrams} (top left).  It carries most of the beam energy, so that its decay products (e.g., an electron-positron pair) are boosted in the forward direction.  The invariant mass of the decay products equals $\mdp$, allowing the dark photon signature to be distinguished from the (large) Quantum Electrodynamics (QED) background through a ``bump hunt'' search in the invariant mass spectrum. This is the strategy pursued by, e.g., APEX~\cite{Essig:2010xa,APEX:2011dww}, the A1 collaboration at MAMI~\cite{A1:2011yso}, HPS~\cite{HPS:2018xkw,Adrian:2022nkt}, DarkLight~\cite{Freytsis:2009bh}, and NA64~\cite{NA64:2018lsq}. 
In addition, for slightly smaller values of $\epsilon$, the dark photon decay length is macroscopic, which can lead to an electron-positron pair emerging from a vertex that is displaced from the electron-beam-target interaction point (the pair seems to emerge out of ``thin air'').  Searches for such displaced vertices are being pursued by HPS (as well as LHCb, see below). 

\textbf{Electron-beam-dump experiments} are similar to fixed-target experiments, but we here distinguish them as having a thick target (see Fig.~\ref{fig:production-A'-schematic} (top right)).  In addition, these experiments often have a thick shield behind the dump and in front of the detector, so that the dark photon has to travel a long distance before reaching (and decaying) in the detector to a fermion-antifermion pair.  Beam-dump experiments are thus excellent at probing long-lived dark photons, and hence smaller values of $\epsilon$~\cite{Bjorken:2009mm}.  Often, they have low backgrounds, since the detector is well-shielded from beam-induced backgrounds. Examples include E137~\cite{Bjorken:1988as}, E141~\cite{Riordan:1987aw}, and E774~\cite{Bross:1989mp} (see~\cite{Bjorken:2009mm,Andreas:2012mt}). 

\textbf{Proton-beam dumps}, in which a beam of protons hits a target, produce many mesons (see Fig.~\ref{fig:production-A'-schematic} (bottom left)).  Whenever mesons have decay modes that involve a photon, there are also corresponding decay modes that involves a dark photon~\cite{Batell:2009di}.  These dark photons can decay in the detector located downstream of the target.  An example is LSND~\cite{Batell:2009di,Essig:2010gu}. 

\begin{figure}[t!]
    \centering
    \begin{minipage}{0.9\linewidth}
        \centering
        \includegraphics[width=0.9\linewidth]{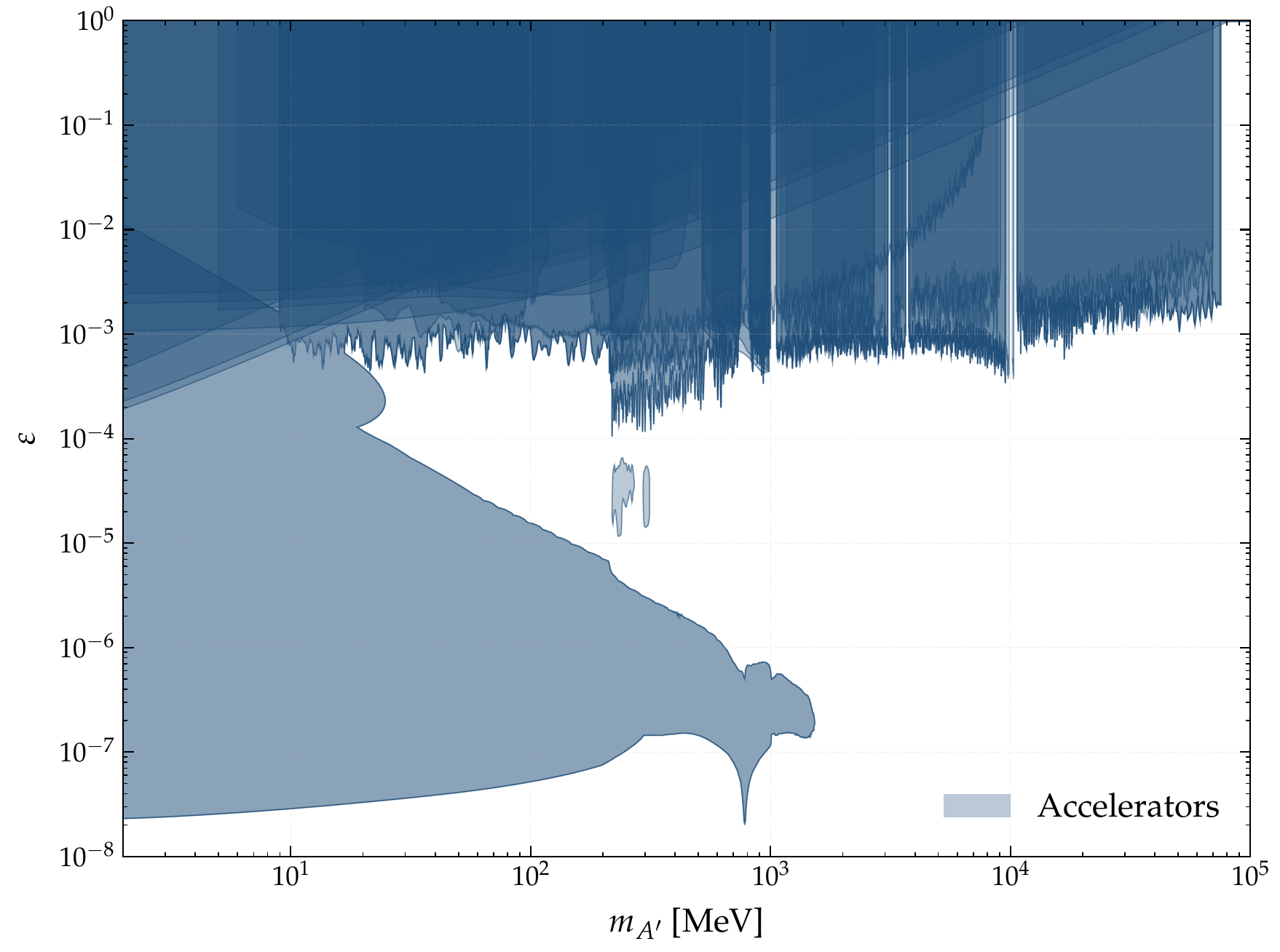}
        \caption{Collider constraints recasted using the \texttt{DarkCast} tool~\cite{DarkCast, Ilten:2018crw} and Ref.~\cite{Kyselov:2024dmi}.\label{fig:ColliderBoundsAboveMeV}}
    \end{minipage}
\end{figure}

\textbf{Electron-positron colliders} (see Fig.~\ref{fig:production-A'-schematic} (bottom right)) can produce dark photons from the diagram shown in Fig.~\ref{fig:production-diagrams} (right) ~\cite{Fayet:2007ua,Batell:2009yf,Essig:2009nc}.  A bump-hunt can again be employed to search for the dark photon. The production cross section is 
\begin{equation}
    \sigma_{\gamma A'} \sim 2.4~{\rm fb} \left(\frac{\epsilon}{10^{-3}}\right)^2 \left(\frac{10.58~{\rm GeV}}{E_{\rm beam}}\right)^2\,,
\end{equation}
where we ignore a possible $t$-channel singularity enhancement when the photon is produced in the forward/backward direction. This shows that the production cross section is larger for smaller beam energies, implying, e.g., that the production cross section at KLOE (with a beam energy near 1~GeV) is larger than at BaBar, Belle, or Belle-II (with a beam energy of around 10~GeV), which is in turn larger than the production of dark photons at the Large Electron-Positron (LEP) collider. Of course, the final sensitivity depends also on the integrated luminosity (and other factors). Decays to, e.g., muon-antimuon pairs will suffer from fewer background events than decays to electron-positron pairs, but both have been done at, e.g., BaBar~\cite{BaBar:2014zli}. 

We note that the dark sector may contain additional particles; in particular, a massive dark photon can be accompanied by a dark Higgs boson, $h'$.  In this case, additional production mechanisms to those discussed above are possible, and more complicated signatures can be produced, see, e.g.,~\cite{Batell:2009yf,Essig:2009nc}.  This includes $A'$-strahlung ($e^+e^- \to A' h'$ where many final states are possible.  For example, if the dark-photon mass is larger than twice the dark Higgs mass, the prompt decay $h'\to A'A'$ is possible, while if the dark-photon mass is smaller, the $h'$ can be long-lived, decaying via, e.g., $h'\to A'A'^* \to A'e^+e^-$, which can produce displaced vertices or no decay in the detector.  Searches for such dark sectors have been done at, e.g., BaBar~\cite{BaBar:2012bkw}, Belle~\cite{Jaegle:2015fme}, KLOE-2~\cite{KLOE-2:2015nli}, and Belle-II~\cite{Belle-II:2022jyy}. 
In addition to a dark Higgs, the dark sector may also contain other states, such as dark matter, and a non-Abelian dark gauge group that, in analogy to Quantum Chromodynamics, can imply the existence of multiple dark gauge bosons.  The signatures of such dark sectors are rich in possibilities.  We will discuss the presence of dark matter further below, but will not discuss the possibility of more complicated dark sectors. 

\textbf{Proton-proton collisions} at the LHC and future high-energy colliders can produce dark photons in, e.g., Drell-Yan collisions or in exotic decays of the 125~GeV SM Higgs boson~\cite{Curtin:2014cca}. They have the best sensitivity to dark photon masses above 10~GeV, where the B-factories loose sensitivity. More complicated dark sectors, such as those with non-abelian gauge symmetries, can again lead to a rich array of signatures~\cite{Strassler:2006im,Baumgart:2009tn}. 
In addition, LHCb is very sensitive to dark photons in a mass range that overlaps the fixed-target searches mentioned above, including from meson production and decay~\cite{Ilten:2015hya,Ilten:2016tkc}. 

\textbf{Rare meson decays} can be used to probe dark photons~\cite{Pospelov:2008zw,Reece:2009un}.  For example, the dark photon can be produced in the decays $\pi^0\to\gamma A'$, $\eta\to\gamma A'$, $\omega,\rho^0,\phi\to\pi^0 A'$, and $K^+\to\pi^+ A'$.  Searches for rare decay modes have not reached the sensitivity of the other searches discussed above, but these decays are often the dominant production mechanisms for dark photons, especially at, e.g., proton beam-dumps. 

A summary of the accelerator-based bounds is shown in Fig.~\ref{fig:ColliderBoundsAboveMeV}. 
\begin{figure}[t]
    \centering
    \includegraphics[width=0.8\linewidth]{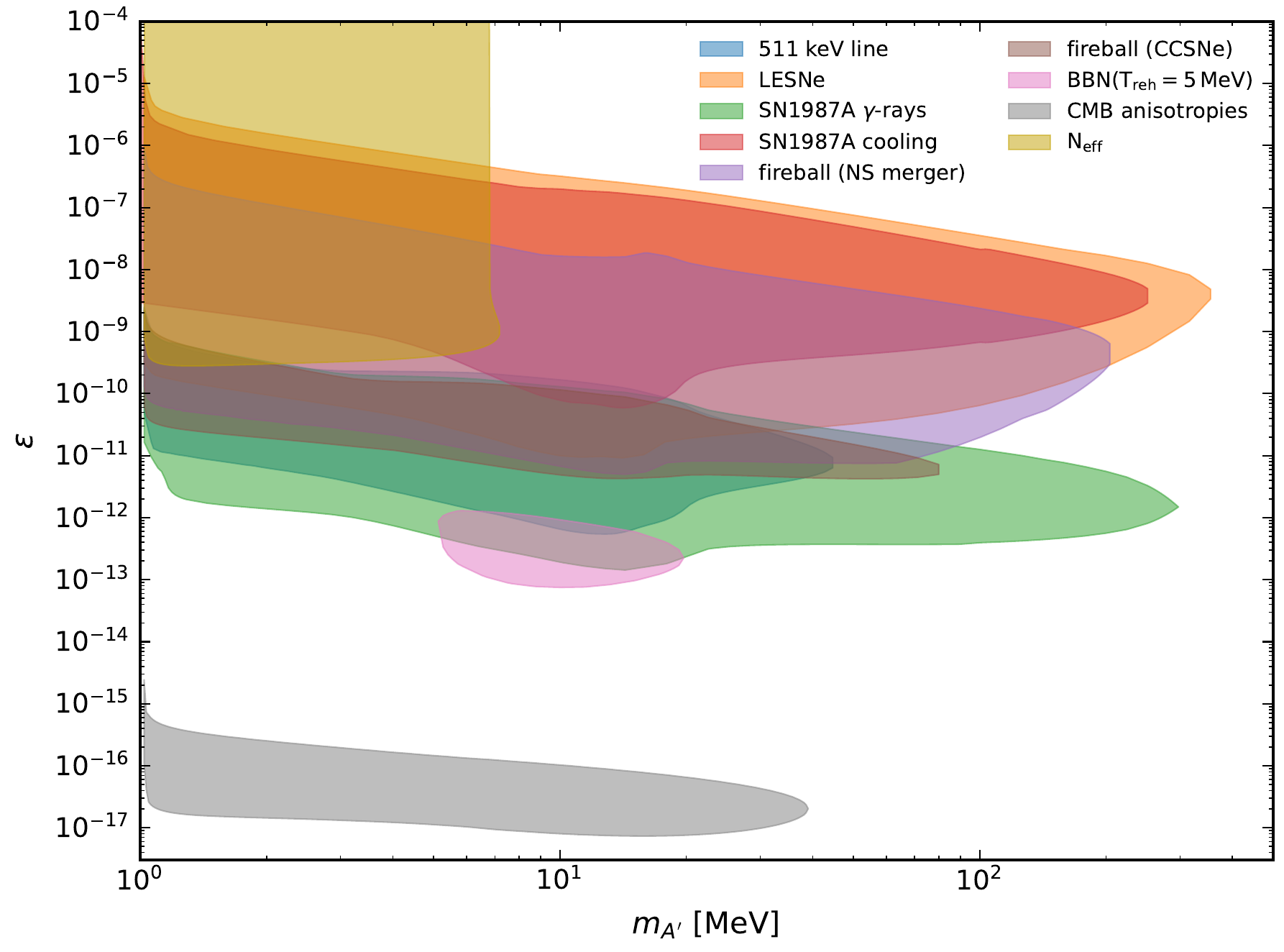}
    \caption{Bounds on massive dark photons from SN-related probes and cosmological limits from BBN and CMB. \label{fig:AstroOnlyAboveMeV}}
\end{figure}

\subsection{Astrophysical and cosmological signatures}

In this subsection, we discuss the astrophysical and cosmological bounds on dark photons. We summarize these bounds in Fig.~\ref{fig:AstroOnlyAboveMeV}; one can find a more exhaustive discussion in~\cite{Caputo:2025avc}. 

\subsubsection{Core collapse supernovae}\label{subsec:SN}

\begin{figure}[t!]
\centering
\includegraphics[width=.4\linewidth]{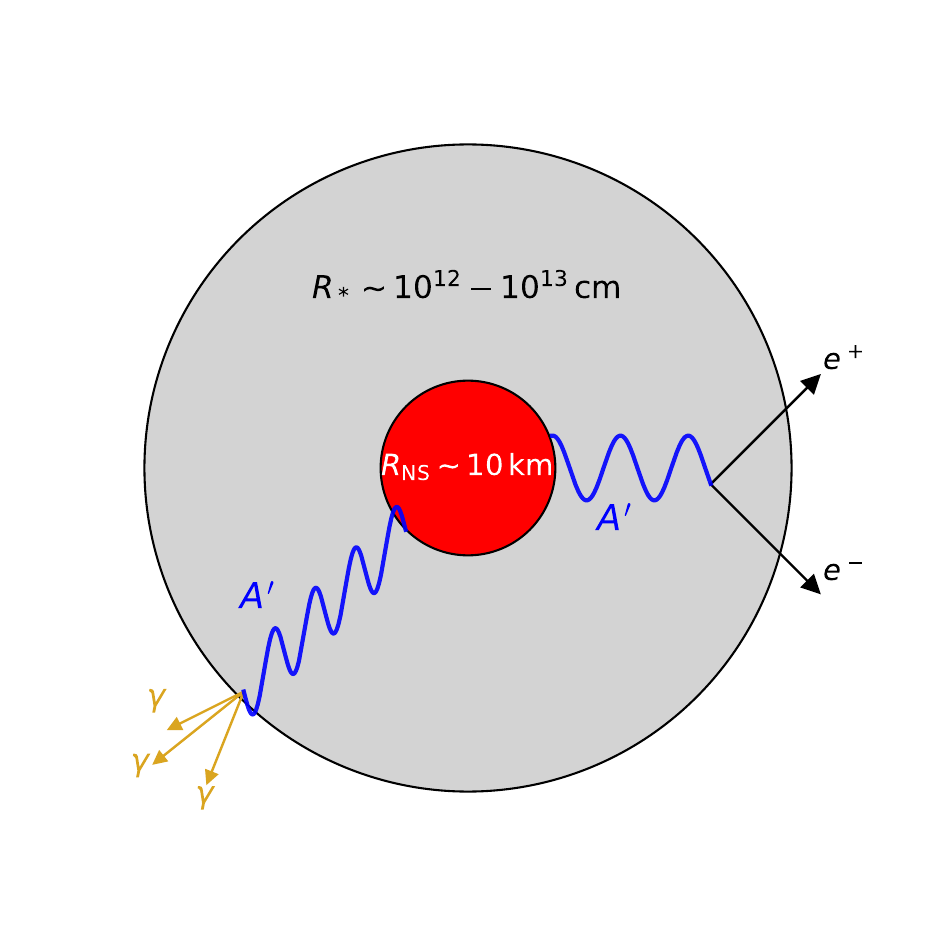}
 \caption{Dark Photons can be copiously produced in the inner region of a CCSN, where the material is very hot and dense, and then decay on their way out into electron-positron pairs,  if $m_{A'} > 2 m_e$, and three photons. The rate of this latter process scales very fast with $\mdp$, becoming more important for large masses.} 
 \label{fig:DecayInSN}
\end{figure}

Some other very important probes of dark photons rely on their production during the collapse of massive stars (those with masses above $\sim 8 M_\odot$), which end their lives in a spectacular explosion known as a Core-Collapse Supernova (CCSN)~\cite{Janka:2012wk, Janka2017Handbooka, Burrows:2020qrp, Boccioli:2024abp}. During this phase, the central engine of the Supernova (SN) reaches incredibly high temperatures, $T \sim 30$–$50$ MeV, and nuclear densities, over a radial extension of $R_{\rm core} \sim 10$ km—comparable to the size of a neutron star, the typical remnant of the explosion. These extreme conditions constitute a powerful locus for the production of very feebly interacting particles, which can be generated and then escape from the central regions of the exploding star~\cite{Raffelt:1996wa, Caputo:2024oqc}. This is the case, for example, for neutrinos, and it may also be the case for dark photons (see Fig.~\ref{fig:DecayInSN}).

The production of dark photons depends on the kinetic-mixing parameter $\epsilon$ and the dark photon mass, as well as the volume of the central core and its properties (temperature and density).  It can be approximated as~\cite{Chang:2016ntp, An:2013yfc, Redondo:2013lna} 
\begin{equation}
    L_{A'} \sim V_{\rm core}\frac{\varepsilon^2 m_{\gamma^\prime}^4}{2\pi^2 } \int d\omega \frac{\omega^2 v_{\gamma^\prime}}{e^{\omega/T}-1} \times \sum_{i={\rm T,L}} \frac{g_i\left|{\rm Im}\,\pi_i\right|}{\left(m_{\gamma'}^2 - {\rm Re}\,\pi_i\right)^2+\left|{\rm Im}\,\pi_i\right|^2}\ ,
\end{equation}
where $ V_{\rm core}$ is the rough volume of the central dense and hot region, $v_{\gamma'}=(1-m_{\gamma^\prime}^2/\omega^2)^{1/2}$, $g_{\rm T} = 2$ for the transverse and $g_{\rm L}=1$ for the longitudinal polarization, and  $\pi_{\rm T, L}$ are the transverse and longitudinal projections of the photon polarization tensor.

Once the dark photons are produced in the central region of the star, they can give rise to a variety of signatures, depending on their mass and kinetic mixing. The most studied and well known constraint comes from the duration of the neutrino signal from SN1987A. In fact, if $A'$ were produced efficiently during the CCSN, then the central engine would have cooled down faster than expected, and the corresponding neutrino cooling signal should have been shorter than what observed. This is at the basis of the so-called ``Raffelt criterion,'' which requires the luminosity into $A'$s to be smaller than the one into neutrinos $L_{A'} \lesssim L_\nu$ at around one second post-bounce (post-bounce refers to the phase after the stellar core has collapsed and rebounded (or "bounced") due to the stiffening of nuclear matter at nuclear densities).  

For $\mdp\gtrsim$1~MeV, the phenomenology is rather interesting and rich compared to smaller $\mdp$, and a variety of probes arise since the $A'$s can decay into electron-positron pairs.  The decay rate is given by
\begin{equation}
\Gamma(A' \to e^+ e^-) \simeq 3.7 \times 10^{-3} \, \text{s}^{-1} \left(\frac{\epsilon}{10^{-11}}\Big)^2 \right(\frac{m_{A'}}{10 \, \text{MeV}}\Big)\ ,
\end{equation}
such that the mean free path for $A'$s produced during the CCSN with an energy $\omega$ will be (taking into account the Lorentz boost factor)
\begin{equation}
    \lambda_{A' \to e^+ e^-}= \frac{1}{\Gamma(A' \to e^+ e^-)}\frac{\omega}{m_{A'}} \simeq 4 \times 10^{13} \text{cm}\, \left(\frac{10 \text{MeV}}{m_{A'}}\right)^2 \left(\frac{10^{-11}}{\varepsilon}\right)^2 \left(\frac{\omega}{50 \, \text{MeV}}\right)\ .
\end{equation}

Depending on the value of the mean free path, different constraints may apply:
\begin{itemize}
    \item If the mean free path is larger than the SN core, but smaller than the progenitor radius, $R_* \sim 10^{12}-10^{13}$ cm, then electromagnetic energy is deposited and quickly absorbed in the stellar envelope. This energy injection is severely limited by electromagnetic observations of CCSNe, in particular low energy SNe~\cite{Caputo:2022mah}.
    \item If the mean free path of the dark photon ($A'$) exceeds the size of the progenitor star, $A'$s produced in the stellar core can escape into outer space and decay into electron-positron pairs. The resulting positrons interact with ambient electrons through Coulomb scattering, leading to radiation and eventual annihilation. The annihilation of non-relativistic positrons contributes, for instance, to the observed 511 keV $\gamma$-ray line in our galaxy. Consequently, all supernovae (SNe) in the Milky Way would produce a significant number of $A'$s, which decay into $e^+e^-$ pairs and contribute to the 511 keV $\gamma$-ray flux; the fact that this latter has been precisely measured by the SPI spectrometer aboard the INTEGRAL satellite then leads to severe constraints~\cite{Balaji:2025alr, DelaTorreLuque:2024zsr}.
    \item For mean free paths somewhat larger than the progenitor radius, a fireball can also form, giving rise to a flux of X-ray photons~\cite{Diamond:2021ekg, Diamond:2023scc}. 
\end{itemize}
\vspace{0.1cm}

The decay into electron-positron pairs is not the only kinematically allowed channel for the dark photon ($A'$). It can also decay into \textit{three SM photons}. This is consistent with the Landau–Yang theorem, which forbids a spin-1 particle, such as the dark photon, or also the $\rho$ meson or the $Z$ boson in the SM, from decaying into two photons. This decay channel is radiatively induced by an electric-charged particle loops, and its rate in the rest frame of the dark photon reads~\cite{McDermott:2017qcg, Pospelov:2008jk}
\begin{equation}
\Gamma_{3\gamma} = \varepsilon^2 \frac{17 \alpha_{\rm em}^4}{2^7 3^6 5^3 \pi^3}\frac{m_{\gamma^\prime}^9}{m_e^8} \mathcal{G}\left(m_{\gamma^\prime}\right)\, \simeq 4.3 \times 10^{-6} \text{s}^{-1} \Big(\frac{\epsilon}{10^{-11}}\Big)^2 \Big(\frac{m_{A'}}{10 \, \text{MeV}}\Big)^9 \mathcal{G}(m_{\gamma^\prime})\ ,
\label{eq:DPdecay3gamma}
\end{equation}
where $\mathcal{G}(m_{\gamma^\prime})$ function quantifies the correction to the dark photon decay rate based on the Euler-Heisenberg action. 
We thus see that for large $\mdp$ actually this decay channel can easily become the dominant one, giving rise to further constraints, especially when the mean free path is larger than the stellar progenitor size. In this case, the $A'$s produced during the core-collapse give rise to a large flux of gamma rays, which is constrained by the Gamma-Ray Spectrometer (GRS) aboard the Solar Maximum Mission satellite in the case of the specific SN1987A event, and by other instruments, such as COMPTEL and Fermi LAT, for the diffuse flux originating from the population of all past CCSNe~\cite{DeRocco:2019njg}. 

\subsubsection{Big Bang Nucleosynthesis and Cosmic Microwave Background}
\vspace{0.1cm}

\begin{figure}[t!]
\centering
\includegraphics[width=.8\linewidth]{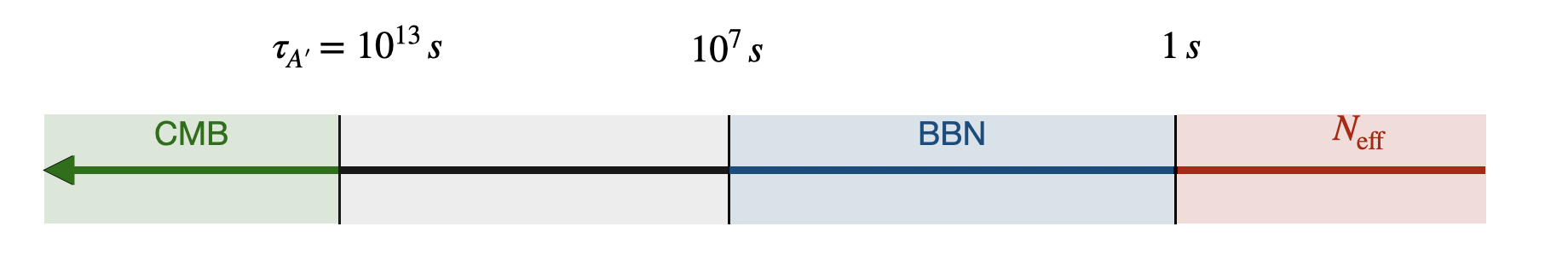}
 \caption{Depending on the $A'$ lifetime, $\tau_{A'}$, different cosmological observables can be impacted. For  lifetimes shorter than $\sim 1$s, the effective number of relativistic species, $N_{\rm eff}$, is the main cosmological probe. More precisely, $A'$ with longer lifetime can still impact $N_{\rm eff}$, but typically the corresponding energy density is too small to leave detectable imprints in $N_{\rm eff}$.
 For larger lifetimes, $ 1 \lesssim \tau_{A'} \lesssim 10^{7}$ s, the abundance of light elements produced during BBN is affected. Finally, for lifetimes larger than the recombination time, $\tau_{A'} \sim 10^{13}$ s, CMB spectral distortions become relevant.} 
 \label{fig:CosmoDP}
\end{figure}

Similarly to the hot and dense material in the central engine of a CCSN, the plasma in the early universe also serves as a natural source for the production of massive dark photons, opening up the possibility to use cosmology to discover, or constrain, the existence of this new particle. In fact, even if these thermally produced dark photons contribute only a small fraction of the total dark matter density, their decay into the electromagnetic sector alter Cosmic Microwave Background (CMB) and Big Bang nucleosynthesis (BBN) measurements. The production of dark photons from the electron-photon plasma in the early Universe depends only weakly on the reheating temperature. This is because, being the kinetic mixing a dimension-4 operator, $A'$ production is infrared-dominated and thus peaks around $T \sim m_{A'}$ (assuming $m_{A'} > m_e$). As a result, the electromagnetic energy released per baryon can be parametrically written as~\cite{Fradette:2014sza}
\begin{equation}
    E_{A' \rightarrow \text{e.m.}} \sim \frac{m_{A'} \, (d\Gamma_{\rm prod}/dV) \, H^{-1}_{T = m_{A'}}}{n_b(T = m_{A'})}\ ,
\end{equation}
where $d\Gamma_{\rm prod}/dV$ is the $A'$ production rate per unit volume, $H$ is the Hubble parameter (so $H^{-1}_{T = m_{A'}}$ is the Hubble time over which $A'$ production is active), and $n_b$ is the baryon number density, all evaluated at $T = m_{A'}$.

To obtain an order of magnitude estimate, we make a few simplifying assumptions, neglecting $\mathcal{O}(1)$ numerical factors. We consider a radiation-dominated Universe, where $H \sim T^2 / M_{\rm Pl}$, with $M_{\rm Pl} = 1.22 \times 10^{19} \, \text{GeV}$ the Planck mass. We approximate the production rate as $d\Gamma_{\rm prod}/dV \sim n_{\gamma, e^\pm} \, \tau_{A'}^{-1}$, where $\tau_{A'}$ is the $A'$ lifetime and $n_{\gamma, e^\pm}$ is the number density of the SM particles sourcing the $A'$s. We also express the baryon number density as $n_b = n_{\gamma, e^\pm} \, \eta_b$, with $\eta_b$ the baryon-to-photon ratio. Putting this together, we find
\begin{equation}
    E_{A' \rightarrow \text{e.m.}} \sim \text{MeV} \, \left( \frac{\epsilon}{10^{-14}} \right)^2.
\end{equation}

The large Planck mass and the small $\eta_b$ can easily compensate for very small values of the kinetic mixing $\epsilon$. Since Big Bang Nucleosynthesis (BBN) is sensitive to energy injections as low as $\mathcal{O}(\text{MeV})$ per baryon, and the cosmic microwave background (CMB) anisotropy spectrum can probe sub-eV energy injections, these cosmological observables provide extremely sensitive probes of such scenarios. Which probe is impacted depends of course on the $A'$ lifetime, compared to different important moment in the cosmological history of our universe, namely neutrino decoupling, the end of BBN, and recombination (see Fig.~\ref{fig:CosmoDP}).

First of all, if dark photons decay after neutrino decoupling (at temperatures around  $T \sim \mathrm{MeV}$, when the universe is $\sim 1\, $s old), the energy they release is deposited solely into the electron-photon plasma, without affecting the neutrino background. This selective heating increases the temperature of the photon bath relative to that of the neutrinos, effectively reducing the neutrino contribution to the total radiation energy density. As a result, the expansion rate of the Universe---when expressed in terms of the photon temperature---is modified. This change alters the evolution of cosmological perturbations, leaving a distinct imprint on precision observables such as the high-\( \ell \) (small-scale) anisotropies in the cosmic microwave background (CMB). The impact of such late decays is commonly captured by a shift in the effective number of relativistic species, $N_{\rm eff}$, a parameter tightly constrained by current CMB measurements~\cite{Ibe:2019gpv}.

If the injection happens on timescales $ 1 \lesssim \tau_{A'} \lesssim 10^{7}$ s, the large number of photons created gives rise to a non-equilibrium destruction and creation of light elements. If $m_{A'} < 2 m_\pi$, the injection is purely electromagnetic (mainly to photons, electrons, or muons, depending on the $A'$ mass), while for $m_{A'} > 2 m_\pi$ the hadronic channels open up~\cite{Fradette:2014sza}. 

Finally, if the injection of electromagnetic energy happens after recombination, relevant for $\tau_{A'} \gtrsim 10^{13}$ s, this would heat up and ionize the neutral gas in the universe.
This leaves observable imprints on the CMB, particularly in the temperature and polarization anisotropy spectra~\cite{Langhoff:2022bij, Bolliet:2020ofj, Slatyer:2016qyl}.

\subsection{Dark photon as a mediator to dark matter}\label{subsec:DPasmediatortoDM}

An idea that has received much attention is that dark matter may be coupled to a dark photon, which is kinetically mixed with the hypercharge gauge boson.  This occurs if the dark matter particle, denoted here as $\chi$, is charged under the $U(1)_D$ gauge symmetry.  This idea is theoretically appealing, as it is simple and potentially has many interesting consequences for a wide range of phenomena, including cosmology, astrophysics, accelerator-based searches, and direct-detection experiments. Such a model is also a useful stand-in for more complicated dark-sector models. We will discuss searches for this model and for a few of its variations.

\subsubsection{Benchmark models from dark matter production in early Universe}

\begin{figure}[t!]
\centering
\includegraphics[width=.25\linewidth]{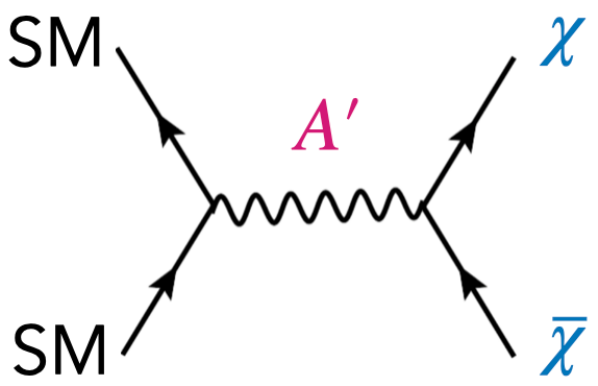}
\caption{Dark matter can be produced in the Universe through ``thermal freeze-out'' with SM particles.  
 } 
 \label{fig:DM-thermal-freeze-out}
\end{figure}

The dark matter can be produced in the early Universe from the annihilation of SM particles.  Various possibilities exist, and for simplicity we focus on the case where $\mdp>2 m_\chi$, as it is the most predictive.  The processes SM+SM$\leftrightarrow \chi\overline\chi$ through an off-shell dark photon, where SM is a SM particle, can be in thermal equilibrium at high temperatures (see Fig.~\ref{fig:DM-thermal-freeze-out}).  As the Universe expands and cools, this process stops and the dark matter and SM sectors decouple, leading to the usual thermal freeze-out scenario~\cite{Boehm:2003hm}.  Neglecting $m_\chi$ and the SM fermion masses, the annihilation cross section scales as 
\begin{equation}\label{eq:DM-freeze-out}
    \sigma v_{\rm rel} \propto \frac{\epsilon^2\alpha_D\alpha m_\chi^2}{\mdp^4}\ v_{\rm rel}^n\,,
\end{equation}
where $v_{\rm rel}$ is the relative velocity between the dark-matter particles.  The exponent $n=0$ if the dark matter is a Dirac fermion (leading to so-called `s-wave' annihilation) and $n=2$ if the dark matter is a complex scalar (`p-wave' annihilation). Requiring that the dark matter relic density corresponds to the observed dark matter abundance fixes the combination of parameter in Eq.~(\ref{eq:DM-freeze-out}).  This then gives specific theoretical benchmark targets that terrestrial experiments can pursue~\cite{Boehm:2003hm,Izaguirre:2015yja,Essig:2015cda,Battaglieri:2017aum}. 
In the simplest model setup, only the complex scalar dark matter is viable, since its late-time annihilation to SM particles 
is p-wave suppressed (i.e., the annihilation rate is suppressed by $v_{\rm rel}^2$); 
in contrast, the Dirac fermion model is not velocity-suppressed, and hence ruled out from, e.g., measurements of the CMB~\cite{Madhavacheril:2013cna} and from gamma-ray observations of the Milky-Way galaxy~\cite{Essig:2013goa}. 

Simple model variations can provide other viable benchmark targets in addition to the complex scalar model.  For example, the Dirac fermion model can be made viable if the abundance is set not by thermal freeze-out, but by an initial asymmetry between the $\chi$ and $\overline\chi$ particles~\cite{Lin:2011gj}.  In addition, instead of a Dirac fermion, one can consider ``pseudo-Dirac'' states, which consists of two Majorana states $\chi_1$ and $\chi_2$ that have a small mass splitting (often called ``inelastic'' dark matter~\cite{Tucker-Smith:2001myb}) (if the two states had the same mass, they would join into a Dirac fermion).  In this case, both $\chi_1$ $\chi_2$ would exist after thermal freeze-out from SM+SM$\leftrightarrow\chi_1 \chi_2$ processes, but the heavier state can quickly decay to $\chi_1$ as $\chi_2\to\chi_1 A'$, where $A'$ can be on-shell or off-shell (in the latter case, the decay is three-body to $\chi_1$ and a SM fermion-anti-fermion pair); there is thus no late-time $\chi_1\chi_2$ annihilation~\cite{Izaguirre:2015yja}.  Accelerator-based experiments are able to probe these various model variations.  In particular, since the dark matter is produced relativistically, even inelastic dark matter can be produced and detected~\cite{Izaguirre:2015yja,Izaguirre:2015zva}.  In contrast, direct-detection experiments are not a good probe of inelastic dark matter, since the dark matter in the Milky-Way halo is non-relativistic. A detailed discussion of this model and its consequences for cosmology cosmology, astrophysics, and colliders can be found in~\cite{Roy:2026sek}.

\begin{figure}[t!]
\centering
\includegraphics[width=.35\linewidth]{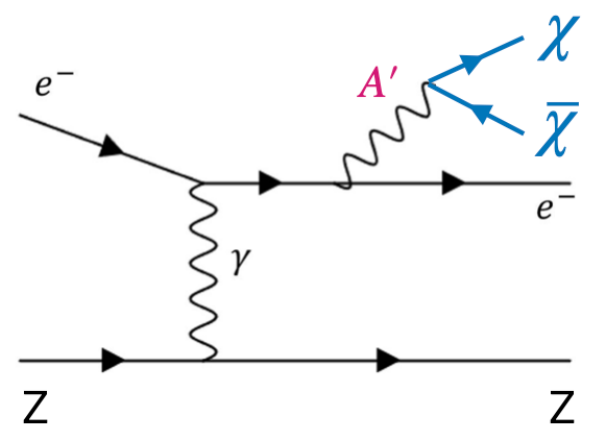}
\hskip 2cm
\includegraphics[width=.2\linewidth]{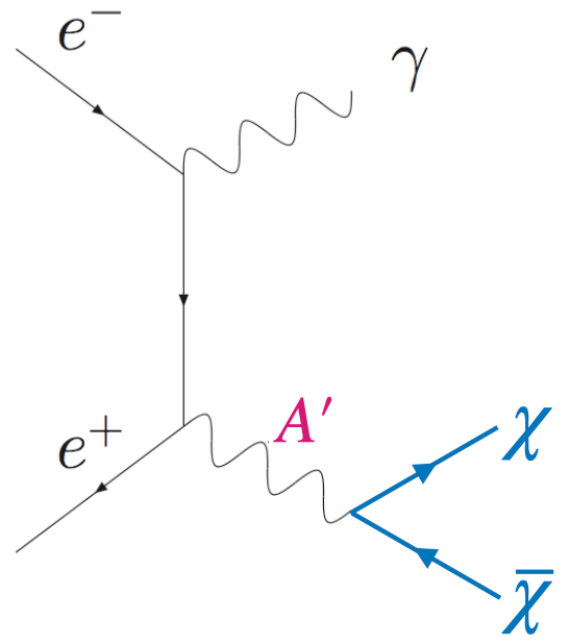}
 \caption{Dark matter can be produced through dark photon decay at electron-positron colliders (left) and at fixed-target experiments (right).} 
 \label{fig:production-diagrams-DM}
\end{figure}

Another important production mechanism is ``freeze-in''~\cite{Hall:2009bx}. In this case, one often assumes that the dark sector is initially empty in the early Universe and get populated by the annihilation of SM particles.  For the dark photon case, the diagram in Fig.~\ref{fig:DM-thermal-freeze-out} would still be applicable, but the couplings are very small to avoid the two sectors being in thermal equilibrium.  Instead, the observed relic abundance builds up over time.  For heavy dark photon mediators, the freeze-in production provides benchmark targets that are out of reach of all proposed terrestrial experiments~\cite{Chang:2019xva}.  However, for ultralight mediators ($\mdp\ll \mathcal{O}(\textrm{keV})$), the freeze-in target~\cite{Essig:2011nj,Chu:2011be,Essig:2015cda,Dvorkin:2019zdi} is within reach of low-threshold direct-detection experiments~\cite{Essig:2011nj,Knapen:2017ekk,Essig:2022dfa,Emken:2024nox}, since the direct-detection scattering cross section scales as $\sigma\propto 1/q^4$, where $q$ is the momentum transfer.  In contrast, accelerator-based experiments are unable to probe this freeze-in target. We will briefly revisit this scenario in \S\ref{subsec:DMwithultralightDPmediator}.

\subsubsection{Accelerator-based probes}

\begin{figure}[t!]
\centering
\includegraphics[width=.38\linewidth]{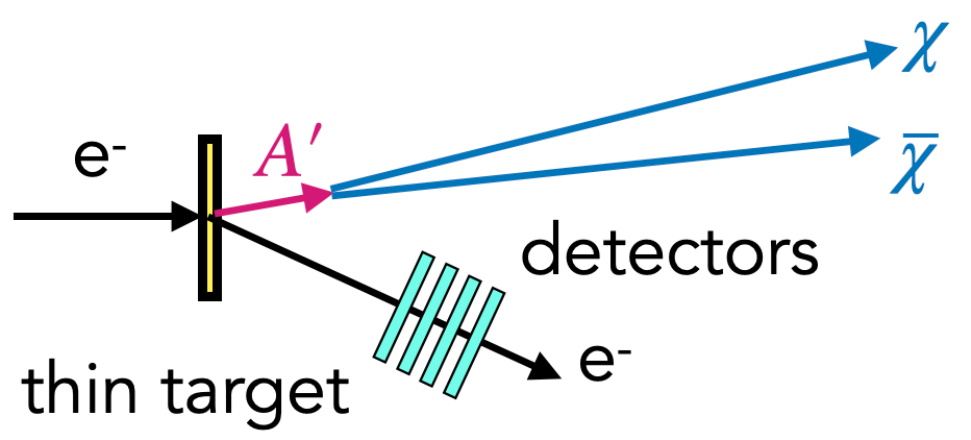}
~~~~~~~~
\includegraphics[width=.49\linewidth]{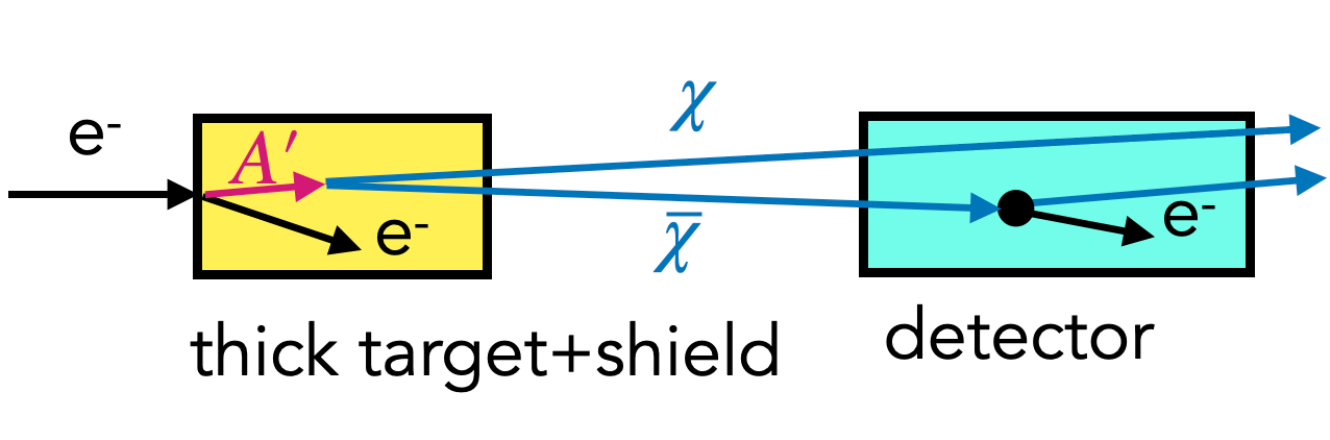}\\
\vskip 5mm
\includegraphics[width=.48\linewidth]{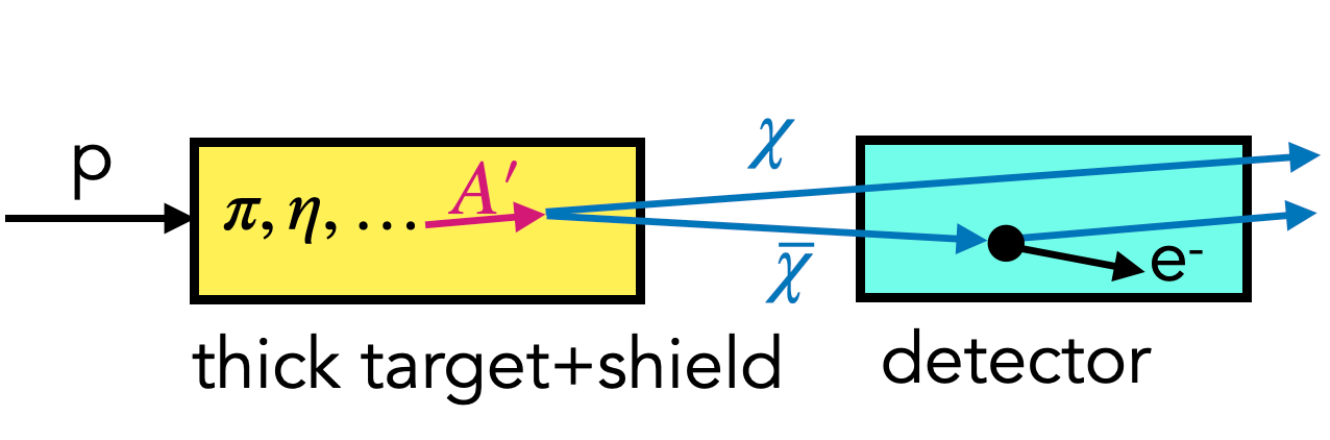}
~~~
\includegraphics[width=.45\linewidth]{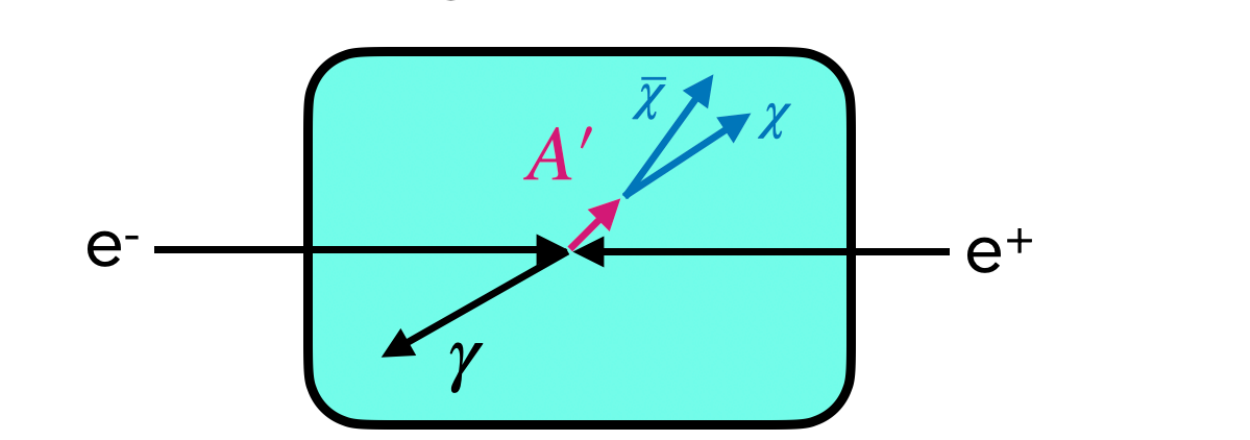}
 \caption{Dark matter can be produced via a dark photon mediator at electron fixed-target experiments (top left), electron beam dump experiments (top right), proton beam-dump experiments (bottom left), and electron-positron colliders.} 
 \label{fig:production-DM-schematic}
\end{figure}

Accelerator-based experiments provide a strong probe of dark matter coupled to a dark photon.  We briefly discuss several examples. 

\textbf{Electron-beam fixed-target experiments.}
The dark photon can be radiated from an electron beam striking a fixed target and subsequently, if $\mdp>2m_\chi$, decay to dark-matter particles (see Figs.~\ref{fig:production-diagrams-DM} and~\ref{fig:production-DM-schematic} top left). The dark matter does not interact with the detector after it is produced (it is ``invisible''), but the energy and momentum of the incoming electron will be altered as it recoils.  By measuring the energy of the recoiling electron, one would infer that there is ``missing'' energy~\cite{Gninenko:2013rka,Andreas:2013lya}. 
Alternatively, by measuring the momentum of the recoiling electron, one would infer that there is ``missing'' momentum~\cite{Izaguirre:2014bca,Izaguirre:2015yja}. Both would indicate that something invisible was produced, and although much more work would be required to prove that a signal is actually from a dark-matter particle that was produced, it is an excellent probe of the model in which dark matter is coupled to a dark photon. Missing-energy searches have already been performed by NA64~\cite{NA64:2016oww,NA64:2017vtt,NA64:2023wbi}, while the LDMX experiment plans to perform a missing-momentum search~\cite{LDMX:2018cma}. 

\textbf{Electron-beam-dump experiments.}
The dark photon is produced as in fixed-target experiments and can decay to dark matter particles (see Fig.~\ref{fig:production-diagrams-DM} (right)).  The dark matter can then travel through a thick shield before traversing a detector, in which it can scatter off electrons or nuclei in the detector, leaving a tell-tale signature.  Constraints come from re-interpreting data from  SLACmQ~\cite{Diamond:2013oda} and E137~\cite{Batell:2014mga} and a further search is planned with the BDX experiment at Jefferson Lab~\cite{BDX:2014pkr,BDX:2016akw,BDX:2017jub,BDX:2019afh}. 

\textbf{Proton-beam-dump experiments.}
The dark photon is produced in decays of mesons that are produced in the proton dump.  The dark photon subsequently decays to dark matter particles and scatters in a detector downstream after passing through shielding.  Constraints exist from re-interpreting data from LSND~\cite{Batell:2009di} and from recent dedicated searches from MiniBooNE~\cite{MiniBooNE:2017nqe,MiniBooNEDM:2018cxm}. 

\textbf{Electron-positron colliders.}
The dark photon is produced in the process $e^+e^-\to\gamma A'$, with the subsequent decay $A'\to\chi\overline\chi$ appearing as missing energy.  Searches for mono-photon events have been done at BaBar~\cite{Essig:2013vha,BaBar:2017tiz}, with Belle-2 expected to set further constraints~\cite{Essig:2013vha}. 

\subsubsection{Dark matter direct-detection probes}\label{subsubsec:DMDD-heavymediator}

\begin{figure}[t!]
\centering
\includegraphics[height=.25\linewidth]{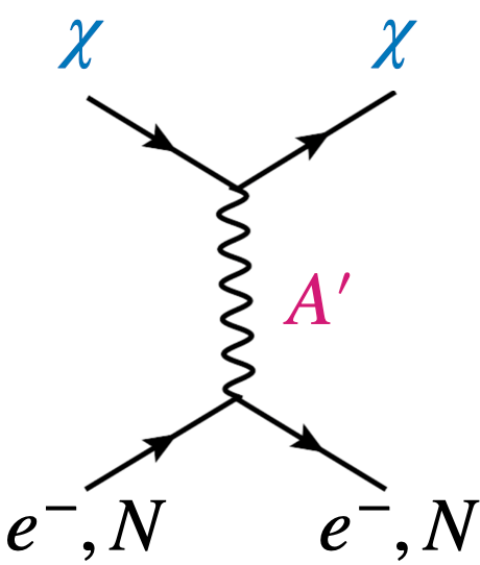}
\hskip 2cm
\includegraphics[height=.25\linewidth]{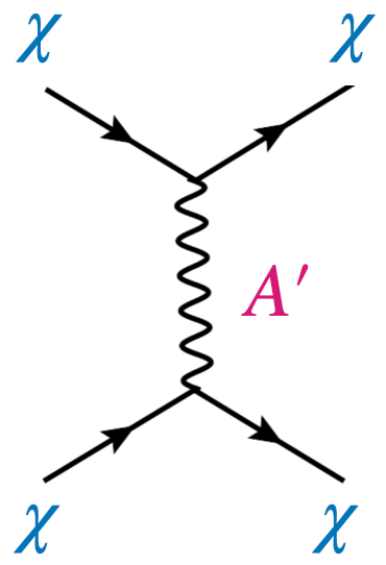}
 \caption{\textbf{Left:}  Dark matter can scatter in off nuclei ($N$) or electrons ($e^-$) in dark matter direct detection experiments.
 \textbf{Right:} Dark matter can self-scatter leading to self-interacting dark matter.} 
 \label{fig:DM-scattering}
\end{figure}

Direct-detection experiments, in which dark matter particles in the Milky-Way halo scatter in a target material (Fig.~\ref{fig:DM-scattering} (left)), are an important complementary probe to accelerator-based searches~\cite{Goodman:1984dc}.  For dark-matter masses above the GeV-scale, the large noble-liquid detectors provide the best constraint on dark matter scattering with nuclei~\cite{LZ:2022lsv,PandaX:2024qfu,XENON:2025vwd}. For lower masses, dark matter-electron scattering sets strong constraints~\cite{Essig:2011nj,SENSEI:2023zdf,DAMIC-M:2025luv}. Several ideas exist to improve the sensitivity to lower interactions and probe various benchmark models~\cite{Essig:2022dfa,Battaglieri:2017aum}.  

\subsubsection{Cosmological and astrophysical probes}

\textbf{Self-interacting dark matter.}
Dark matter can scatter with itself through dark-photon exchange, leading to ``self-interacting'' dark matter (Fig.~\ref{fig:DM-scattering} (right)).  This can impact the formation and evolution of structure on sub-galactic scales~\cite{Feng:2009hw,Buckley:2009in,Loeb:2010gj}. Self interacting dark matter is receiving much attention currently~\cite{Tulin:2017ara}, and this model, or variations thereof, provides one avenue to obtain the desired properties. 

\textbf{Supernova 1987A.}
We discussed in \S\ref{subsec:SN} that SN1987A can constrain dark photons that decay to ordinary matter.  If the dark photons are coupled to dark matter, it is still possible to obtain constraints~\cite{Chang:2018rso}. For dark photons that can decay to dark matter, the bounds are complementary to terrestrial probes, constraining parameter space towards smaller interactions that lies slightly below the sensitivity of near-term direct-detection and accelerator-based experiments. Bounds from SN1987A for other scenarios are discussed in~\cite{Chang:2018rso}.

\begin{figure}[H]
	\centering
	\includegraphics[width=0.8\textwidth]{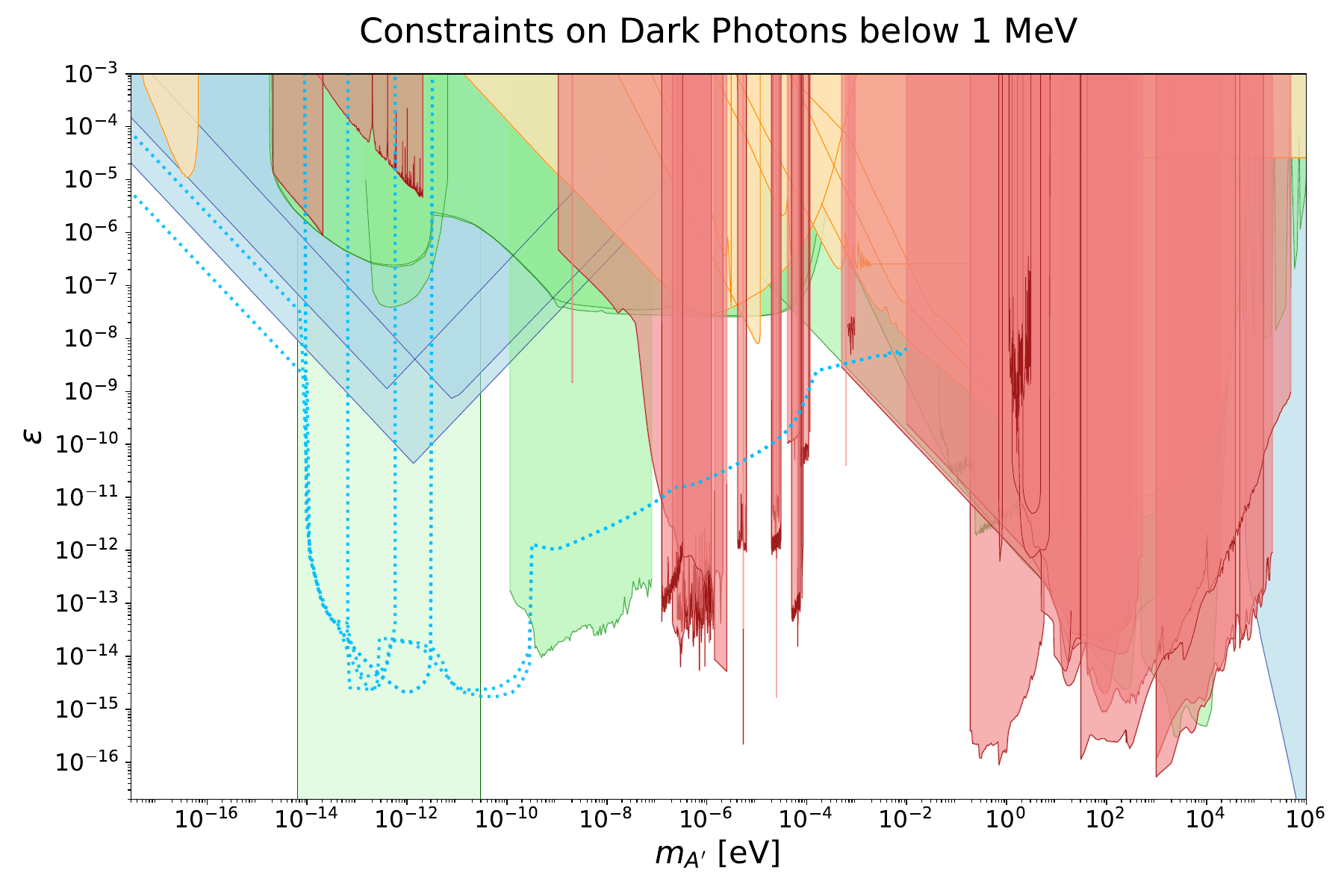}	
\caption{Constraints on dark photons with masses below MeV. Regions in red (light orange) indicate parameter space excluded by laboratory experiments assuming dark photons are (are not) dark matter. Cosmological and astrophysical bounds are shown in blue and green, respectively; the blue limits assume that dark photons constitute all of the dark matter. The dotted lines indicate the bounds that may be affected by the recent argument in Ref.~\cite{Hook:2025pbn}. We extracted all the data from the \texttt{AxionLimits} GitHub repository~\cite{AxionLimits} curated by Ciaran O'Hare (see also Ref.~\cite{Caputo:2021eaa}).}
	\label{fig:BoundsBelowMeV}
\end{figure}

\section{Dark photons with mass below 1~MeV}

In this section, we focus on dark photons with masses below 1~MeV.  A summary of the bounds is shown in Fig.~\ref{fig:BoundsBelowMeV}. 

\subsection{Searches for dark photons irrespective of their connection to dark matter}

\vspace{0.2cm}

In this section we discuss a variety of probes for dark photons with masses below (including much below) the MeV scale, without any assumption about them constituting a part of the dark matter.
In general, collider probes are not the best probes of dark photons in this mass range, although a proposal does exist to use a laser beam (with energy $\sim$1~eV) incident on an electron beam (with energy $\sim$3~GeV) to produce dark photons from inverse Compton scattering, potentially surpassing existing bounds for dark photon masses in the range $3\times 10^{-5}$~eV to $3\times10^{-3}$~eV~\cite{Angel:2026hfp}.

\subsubsection{Searches for non-Coulomb forces}
\vspace{0.1cm}

The Coulomb potential between two unit charges is given by
\begin{equation}
    V(r) = - \frac{e^2}{4\pi r}\ .
\end{equation}
However, if the dark photon exists, we would all have been wrong! In fact, in the presence of a massive dark photon, the Coulomb potential instead takes the form
\begin{equation}
    V(r) = - \frac{e^2}{4\pi r} \left( 1 + \epsilon^2 e^{-\mdp r} \right) \equiv V_{\rm Coulomb} + \delta V_{A'}\ ,
\end{equation}
which includes an additional Yukawa-like correction with a characteristic range set by $\mdp^{-1}$. 

This modification opens up the possibility of searching for dark photons through high-precision measurements of electrostatic forces, which have been carried out for over a century. One such example is the experiment first performed by Plimpton and Lawton in 1936, where the potential difference between two concentric thin metal shells is measured with high accuracy~\cite{Kroff:2020zhp}. A related idea, dating back to Henry Cavendish’s 1773 experiment (though not motivated by dark photons), uses multiple nested conducting shells to probe deviations from Coulomb’s law. These so-called Cavendish-type experiments have also been used to constrain the photon rest mass~\cite{Tu:2005ge}, and offer sensitivity to longer length scales, around $\sim 10\,\mathrm{cm}$, corresponding to smaller dark photon masses. Another avenue involves precise measurements of the Casimir force, for example using a metallized plate and a sphere mounted on an atomic force microscope (AFM)~\cite{Kroff:2020zhp}. 

Finally, atomic transitions provide a complementary probe: the presence of hidden photons would lead to small, first-order shifts in atomic energy levels. For a given atomic state $|\psi_n\rangle$, the leading-order energy shift can be computed using standard perturbation theory as
\begin{equation}
\delta E_n^{(1)} = \langle \psi_n \,|\, \delta V_{A'} \,|\, \psi_n \rangle\ ,
\label{Eq:EnergyShift}
\end{equation}
which is valid for small perturbations, consistent with the absence of any observed large deviations from QED predictions. These shifts have been computed for atomic hydrogen, hydrogen-like ions, and exotic atoms in~\cite{Jaeckel:2010xx}, and have been used to constrain dark photons down to sub-nanometer scales. 

The presence of a massive dark photon can also lead to modifications of magnetic fields~\cite{Kloor:1994xm}. This opens the possibility of using precise satellite measurements of Earth's magnetic field~\cite{Fischbach:1994ir}, as well as Jupiter’s field from the Juno~\cite{Yan:2023kdg} and Pioneer-10~\cite{Davis:1975mn} missions, to search for such effects. In this case, the relevant length scales are much larger, allowing sensitivity to dark photon masses in the range $m_{A'} \sim 10^{-16}$--$10^{-14}~\mathrm{eV}$. For comparison, Jupiter’s radius is $r_{\rm J} \sim 7.1 \times 10^4~\mathrm{km}$, corresponding to an inverse length scale $1/r_{\rm J} \sim 2.8 \times 10^{-15}~\mathrm{eV}$. Interestingly, as early as 1943, Erwin Schrödinger studied the modification of the Earth's and the Sun’s magnetic fields in a theory with massive photons~\cite{Schrodinger1943}.

\subsubsection{Stellar cooling bounds}
\vspace{0.1cm}

If the dark photon mass is around the MeV scale or above, the only stellar engine able to produce these new particles is the core of a CCSN, where temperature and density are high enough, as we already explained. However, for smaller masses, other stars, for which the typical core temperature are around $1-10$ keV, can also be efficient places of dark photon production~\cite{An:2013yfc, Hardy:2016kme, Redondo:2013lna, An:2013yua, An:2020bxd, Alonso-Alvarez:2020cdv}. For dark photon masses above the plasma frequency of the medium, the production from transverse photons dominate, while for smaller dark photon masses longitudinal modes dominate, because the longitudinal mode of the dark photon can then be produced on resonance at any temperature for $m_{A'} \ll \omega_{\rm pl}$. 

In this case also our star, the Sun, becomes a possible powerful source of dark photons. Very strong constraints can then be derived from the requirement that the power emitted from the Sun into dark photons does not affect solar properties; in particular, a combination of helioseismology (sound speed, surface helium and convective radius) and solar neutrino observations imposes~\cite{Vinyoles:2015aba}
\begin{equation}
L_\odot^{A'} \lesssim 0.02\, L_\odot\ ,
\end{equation}
where $L_\odot = 3.83 \times 10^{26}$ Watt is the measured solar luminosity.

Similar constraints, which extend to slightly larger masses, derive from the study of horizontal branch stars. For example, one may require that energy loss into dark photons should not exceed the nuclear energy generation rate in these stars, $ L_{\rm nuc}^{\rm HB} \sim 10^{-5}$ Watt gram$^{-1}$. More sophisticated treatments involve the inclusion of the dark photon into stellar evolution code like the Modules for Experiments in Stellar Astrophysics (MESA). A careful study of this type was recently done in~\cite{Dolan:2023cjs}, where the authors performed stellar evolution simulations, including dark photon cooling channels, and then compared with observations from globular clusters, in particular the red-giant branch (RGB) tip luminosity, and the ratios of RGB to horizontal branch (R-parameter) and asymptotic giant branch to HB stars (R2-parameter).

\subsubsection{CMB spectral distortions}
\vspace{0.1cm}

For masses above the $\sim$MeV scale, we have seen how cosmology provides very strong constraints on the existence of dark photons. In that case, dark photons were thermally produced in the early universe plasma, affecting different observables such as BBN, $N_{\rm eff}$, and CMB. The latter is a very important tool also for much lighter dark photons, $m_{A'} \ll $ eV; in this case, however, production in the early universe is completely irrelevant. So, how do we probe this scenario? Instead of looking for a chain of processes of the type $\gamma, e^{\pm} \rightarrow A' \rightarrow \text{e.m cascade}$, we look for the reverse process of CMB photons disappearances~\cite{Caputo:2020bdy, Caputo:2020rnx, Witte:2020rvb, McDermott:2019lch, Mirizzi:2009iz, Arsenadze:2024ywr} due to kinetic mixing
\begin{equation}
    \gamma_{\rm CMB} \rightarrow A'\ ,
\end{equation}
which generates distortions of the expected black body spectrum. Such a conversion happens with a specific probability, which is particularly enhanced if the dark photon mass, $m_{A'}$, matches the plasma frequency, \textit{i.e}, the effective mass for the photon transverse modes. This latter scales with redshift, tracking the electron density distribution, and is
\begin{equation}
\omega_{\rm pl}^2(z) \simeq 1.4 \times 10^{-21}~\mathrm{eV}^2 \left( \frac{n_e(z)}{\mathrm{cm}^{-3}} \right)\ .
\end{equation}
At the point of resonance, when $m_{A'} = \omega_{\rm pl}(z, \vec{x})$, the conversion probability reads
\begin{equation}
\label{Eq:ConvProb}
P_{\gamma \rightarrow A'} \simeq \frac{\pi \epsilon^2 \omega_{\rm pl}^2(z_{\rm res})}{\omega(z_{\rm res})(1 + z_{\rm res}) H(z_{\rm res})} 
\left| \frac{d \log \omega_{\rm pl}^2}{dz} \right|^{-1}_{\rm res}\ ,
\end{equation}
where everything is evaluated at resonance. Notice that this formula assumes the universe to be homogeneous, a good approximation above $z \sim 20$. However, for redshift below $z \sim 20$, densities fluctuations are large and Eq.~(\ref{Eq:ConvProb}) must be generalized to account for the spectrum of fluctuations~\cite{Caputo:2020bdy, Caputo:2020rnx, Garcia:2020qrp, Witte:2020rvb}. 

In order to gain some intuition about typical values and mass scaling of the conversion probability, let us focus on dark photon masses $m_{A'} \gtrsim 10^{-9}~\mathrm{eV}$, for which conversion occurs deep into the radiation-dominated era. At that time, the hydrogen is fully ionized, and we have $\omega_{\rm pl}^2(z) \propto n_e(z) \propto (1 + z)^3$; also $H(z) = H_0 \, \Omega_r^{1/2} (1 + z)^2$, and thus
\begin{equation}
\text{Radiation era:} \quad P_{\gamma \rightarrow A'}(x) \simeq \frac{\epsilon^2 \mathcal{F}}{x}
\qquad \text{with} \qquad
\mathcal{F} = \frac{\pi m_\gamma^2(z = 0)}{3 \Omega_r^{1/2} H_0 T_0} \simeq 10^{11}\ ,
\tag{5}
\end{equation}
where to compute $\mathcal{F}$ numerically, we used $m_{\rm pl}(z = 0) \simeq 1.7 \times 10^{-14}~\mathrm{eV}$. Then, to estimate a rough bound, we require $P_{\gamma \rightarrow A'} \sim 10^{-4}$, consistent with the COBE-FIRAS sensitivity to blackbody distortion. This yields
\begin{equation}
\epsilon_{\rm est} \sim 3 \times 10^{-8} \left( \frac{P_{\gamma \rightarrow A'}}{10^{-4}} \right)^{1/2}
\left( \frac{10^{11}}{\mathcal{F}} \right)^{1/2}.
\tag{6}
\end{equation}
Of course, this is only a very rough estimate, but it gives an idea of the values of kinetic mixing one can probe.
 
In order to be more precise, one needs to treat differently the conversions that happen in different cosmological eras~\cite{Arsenadze:2024ywr, Chluba:2024wui}. In fact, the specific type of spectral distortions caused by photon-to-dark photon conversion depends critically on when this occurs and on which photon interactions dominate during that epoch. The primary photon-interacting processes are
\begin{align}
\text{Compton Scattering (CS)} &: \quad e^- + \gamma \leftrightarrow e^- + \gamma, \nonumber \\
\text{Double Compton Scattering (DCS)} &: \quad e^- + \gamma \leftrightarrow e^- + \gamma + \gamma, \nonumber \\
\text{Bremsstrahlung (BR)} &: \quad e^- + X \leftrightarrow e^- + X + \gamma\ . \tag{9}
\end{align}

Among these, DCS and BR are photon-number-changing processes. When their interaction rates exceed the Hubble expansion rate, they efficiently drive the photon distribution toward a thermal Bose–Einstein spectrum with vanishing chemical potential. In contrast, CS conserves photon number, but still plays a significant role by redistributing photon momenta and shaping the spectral distribution.

The thermal history of the Universe for $T \lesssim$ keV can be classified into five key regimes, depending on the effectiveness of these processes. These include: the $T$-era ($T \gtrsim 0.5\,\mathrm{keV}$), relevant for $m_{A'} \gtrsim 5 \times 10^{-5}$ eV, the $\mu$-era ($70\,\mathrm{eV} \lesssim T \lesssim 0.5\,\mathrm{keV}$), for $3\times 10^{-6}\lesssim \mdp \lesssim 5 \times 10^{-5}$, the $\mu$-$y$ transition ($2\,\mathrm{eV} \lesssim T \lesssim 70\,\mathrm{eV}$), for $2\times 10^{-8}\lesssim \mdp \lesssim 3\times 10^{-6}$, the $y$-era ($0.2\,\mathrm{eV} \lesssim T \lesssim 2\,\mathrm{eV}$), for  $10^{-10}\lesssim \mdp \lesssim 2\times 10^{-8}$, and the free-streaming regime ($T \lesssim 0.2\,\mathrm{eV}$), relevant for $\mdp \lesssim 10^{-10}$ eV.

\subsubsection{Superradiance}
\vspace{0.1cm}

Dicke introduced the concept of \textit{superradiance} in 1954~\cite{PhysRev.93.99}, describing the collective amplification of radiation resulting from the coherent behavior of emitters. Later, in 1971, Zel’dovich proposed that, under certain conditions, rotating absorptive surfaces can amplify incident radiation—particularly for waves with specific angular properties~\cite{1971JETPL..14..180Z}. This effect, now widely known as (rotational) superradiance, occurs when the frequency $\omega$ of the incoming wave satisfies the condition
\begin{equation}
    \omega < m\Omega,
    \label{Eq:SuperradianceCondition}
\end{equation}
where $m$ is the azimuthal quantum number and $\Omega$ is the angular velocity of the rotating object.

When applied to black holes (BHs), this mechanism becomes the wave analogue of the Penrose process, where particles can extract rotational energy from the black hole. Remarkably, the gravitational potential of a rotating (Kerr) black hole admits quasi-bound states that can trap massive bosonic fields near the black hole. As a result, low-energy bosons that satisfy Eq.~(\ref{Eq:SuperradianceCondition}) undergo repeated superradiant scattering~\cite{Cardoso:2004nk, Brito:2015oca, Arvanitaki:2009fg}. With each interaction, the wave is further amplified, leading to an instability characterized by the exponential growth of these bound states, which form a very dense boson cloud. This process can ultimately spin down the black hole. 

\begin{figure}[t!]
\centering
\includegraphics[width=.5\linewidth]{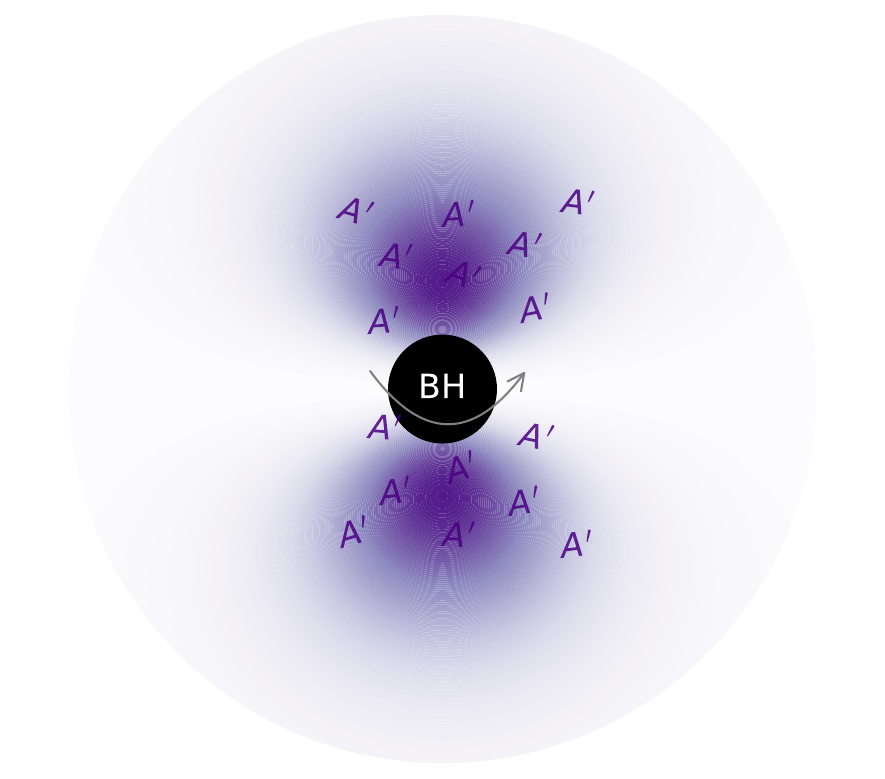}
 \caption{A spinning BH can grow a superradiant cloud of dark photons, with energy densities tens of order of magnitude larger than the local dark matter density.} 
 \label{fig:SupCloud}
\end{figure}

The superradiance phenomenon occurs for any bosonic particle, and therefore also for spin-1 dark photons (see Fig.~\ref{fig:SupCloud}). Superradiance is efficient only if 
\begin{equation}
    m_{A'} \lesssim M^{-1}_{\rm BH}\ ,
\end{equation}
and the cloud formation timescale (for spin-1 particles, and for the fastest growing mode) reads~\cite{Baryakhtar:2017ngi, Cardoso:2018tly, East:2017ovw}
\begin{equation}
\tau_s \simeq \frac{10^2}{\tilde{a}} \left( \frac{M_{\rm BH}}{10 M_\odot} \right) \Big(\frac{0.1}{\alpha}\Big)^7\text{s}\ ,
\end{equation}
where $\tilde{a}$ is the dimensionless spin parameter for a Kerr BH and $\alpha \equiv m_{A'}M_{\rm BH}$. If $\tau_s$ is small compared to other astrophysical timescales, the superradiant cloud will grow until it extracts up to $\sim 10 \%$ of the BH angular momentum. Therefore, the presence of gaps in the BH mass-spin Regge plane, as well as the observation of highly spinning BHs, can be used to constrain the existence of dark photons\cite{Caputo:2025oap, LIGOScientific:2025brd, Aswathi:2025nxa, Cardoso:2018tly, Baryakhtar:2017ngi}. 

Superradiance constraints typically rely only on presence of a massive boson; nevertheless, the presence of interactions with the standard model (for example via the kinetic mixing) can lead to further interesting signatures~\cite{Caputo:2021efm, Cannizzaro:2022xyw, Siemonsen:2022ivj,Berghaus:2025kvn}. 

\subsection{Dark photons as dark matter}

For masses below (or much below) 1 MeV, an appealing possibility is that dark photons can be all or a relevant fraction of the dark matter in our universe. 

\subsubsection{Cosmological production}
\vspace{0.1cm}

A variety of mechanisms have been proposed to generate dark photon dark matter (DPDM) in the early Universe. One of the simplest is the misalignment mechanism, originally developed for axions~\cite{Preskill:1982cy,Abbott:1982af,Dine:1982ah}. For vectors, however, this setup typically requires a non-minimal coupling to the Ricci scalar in order to yield the correct relic abundance~\cite{Arias:2012az,Graham:2015rva,AlonsoAlvarez:2019cgw}, though such couplings may introduce instabilities~\cite{Himmetoglu:2008zp,Himmetoglu:2009qi,Karciauskas:2010as}. Production mechanisms involving  dark freeze-out or dark freeze-in from a frozen-in dark sector are presented in~\cite{Aboubrahim:2021ycj,Feng:2024nkh}. 

An alternative mechanism, which is probably the most elegant and minimal way to produce dark photon, relies on inflationary fluctuations, which can generate a relic population of vectors without the need for additional couplings to gravity~\cite{Graham:2015rva,Kolb:2020fwh,Ema:2019yrd,Ahmed:2020fhc,Nakai:2020cfw}. These models avoid long-wavelength isocurvature constraints by producing a spectrum peaked at intermediate scales, in contrast to scalars and tensors. The relic density in this case is given by
\begin{equation}
\Omega_{A'}
= \Omega_{\text{cdm}}\sqrt{\frac{m}{6 \times 10^{-6}~\mathrm{eV}}}
\left( \frac{H_I}{10^{14}~\mathrm{GeV}} \right)^2\ ,
\end{equation}
where $H_I$ is the Hubble scale during inflation, which the Planck bound on the tensor-to-scalar ratio limits to be $H_I \lesssim 10^{14}$~GeV. Given this upper limit on the Hubble scale, this production mechanism cannot lead to dark photon masses larger than $\sim 10^{-5}$~eV.

Other production channels include tachyonic growth induced by couplings to axions~\cite{Agrawal:2018vin,Co:2018lka,Bastero-Gil:2018uel,Dror:2018pdh}, emission from topological defects such as Abelian-Higgs cosmic strings~\cite{Long:2019lwl,Co:2021rhi, Kitajima:2022lre}, and production from spectator scalar field during inflation~\cite{Nakai:2022dni}.

We notice that the dark photon mass can be generated via both the Higgs and St\"uckelberg mechanisms. In the former case, one should probably re-examine many of the productions mechanisms cited above, as~\cite{East:2022rsi} pointed out that for much of the parameter space of interest vortex formation in the Higgsed phase could impede the efficient production of low momenta dark photon dark matter.  Recent studies in this direction include~\cite{Cyncynates:2023zwj,Cyncynates:2024yxm}.

Most of the mechanisms introduced above naturally produce very light dark photons, with $m_{A'} \ll \text{eV}$ (although some mechanism, such as that in~\cite{Bastero-Gil:2018uel}, can produce also very massive DPDM, up to the TeV scale), and are also non-thermal in nature. On the other hand, the presence of a coupling to electromagnetism implies that dark photons can always be produced from the primordial SM plasma via plasmon resonant conversion or scattering processes~\cite{Redondo:2008ec}. This becomes particularly relevant for masses below $\sim 2\,m_e$, since more massive dark photons decay rapidly into electron–positron pairs, as discussed in previous sections. For lower masses, the dark photon instead decays into three photons, a process whose rate, as noted above, is much smaller and decreases steeply with $\mdp$.  Nevertheless, even for such low-mass dark photons, strong constraints exist on kinetic mixing values relevant for successful freeze-in production. These arise from galactic and extragalactic X-ray observations~\cite{Linden:2024fby, Redondo:2008ec}, as well as from overclosure bounds and direct detection experiments on Earth.

We note that here we do not review the (possibly very rich) phenomenology of dark photons dark matter in the presence of solitons or other substructures, which typically arise purely from the interplay of gravitational interactions and quantum effects~\cite{Gorghetto:2022sue, Schiappacasse:2026ems, Amin:2022pzv, Zhang:2024bjo, Zhang:2021xxa}.

\subsubsection{Dark matter detection in the lab: particle regime}
\vspace{0.1cm}

Typical direct detection experiments for particle dark matter with a mass above, or much above, the eV scale rely on the elastic scattering off ordinary matter of the dark matter candidate; this is the case, for example, for Weakly Interacting Massive Particles (WIMPs). However, if dark matter is made of dark photons, or any other (pseudo)scalar or a vector bosons, then it can actually be \textit{absorbed} by the material, depositing all its energy, kinetic and rest-mass. The net effect is that the DM is absorbed by an atom and an electron is ejected, analogously to the standard photoelectric effect. In fact, the dark photon absorption cross-section can be simply written in terms of the standard photoelectric absorption
cross section, $\sigma_{\rm PE}$, as~\cite{Pospelov:2008jk,Bloch:2016sjj}
\begin{equation}
    \sigma_{A'}(\omega_{A'} = \mdp) \, v_{A'} \simeq \varepsilon^2 \, \sigma_{\rm PE}(\omega = \mdp) \,,
    \label{eq:AbsorptionCrossSection}
\end{equation}
where $v_{A'}$ is the dark matter velocity. The resulting absorption rate is given by
\begin{equation}
    \text{Rate per atom} \simeq \frac{\rho_{\rm DM}}{m_{A'}} \times \epsilon^2 \, \sigma_{\rm PE}(E = m_{A'}) \,,
    \label{eq:AbsorptionRate}
\end{equation}
which can then be convolved with the detector effective exposure $\mathcal{E}(\omega)$ to obtain the expected number of events
\begin{equation}
R_{A'} \simeq 1 \left( \frac{\rho_{\rm DM}}{0.4~\mathrm{GeV}/\mathrm{cm}^3} \right)
    \left( \frac{m_{A'}}{2.5~\mathrm{keV}} \right)
    \left( \frac{\epsilon}{10^{-16}} \right)
    \left( \frac{\mathcal{E}(m_{A'})}{200~\mathrm{ton/day}} \right).
\end{equation}

Different collaborations have been looking for such an absorption process, including SuperCDMS~\cite{Aralis:2019nfa}, XENON~\cite{An:2013yfc,Redondo:2013lna,Aprile:2020tmw,XENON:2021qze}, PANDA-X~\cite{PandaX:2023xgl}, SENSEI~\cite{Crisler:2018gci,SENSEI:2019ibb,SENSEI2020,SENSEI:2023zdf}, DAMIC-M~\cite{DAMIC-M:2025luv}, DarkSide-50~\cite{DarkSide:2022knj}, providing leading constraints for $\mdp \sim 0.1-1$ eV. These direct detection experiments can actually be used to look for dark photons even when they are not dark matter; this is the case for dark photons produced in the Sun, both relativistic~\cite{Redondo:2008aa,An:2013yua,Bloch:2020uzh} and non-relativistic~\cite{Lasenby:2020goo}. 

\subsubsection{Dark matter detection in the lab: wave regime}
\vspace{0.1cm}

For masses below $\sim$eV, the phase-space occupation number for dark matter particles in our galaxy becomes much bigger than one, and it makes sense to treat them as a collection of plane waves, as a classical field, rather than single particles. Then, what matters for typical experiments in this mass range is the (ordinary) electric field produced by the dark photon via the kinetic mixing
\begin{equation}
    \left|\textbf{E}_0 \right| = |\frac{\varepsilon \, m_X}{\epsilon_{\rm medium}}\textbf{A}_0| \,,
\end{equation}
where $\epsilon_{\rm medium}$ is the dielectric constant of the medium (equal to unity in vacuum) and where $\textbf{A}_0$ indicates the spatial components for the zero mode of the dark photon, which can be written in terms of the DM energy density
\begin{equation}
    \rho_{\rm DM} = \frac{\mdp^2}{2}|\textbf{A}'_0|^2 \,,
\end{equation}
implying in vacuum a typical electric field of the size $ \left|\textbf{E}_0 \right| \simeq 3700 \, \frac{\text{V}}{\text{m}} \, \varepsilon \, \Big(\frac{\rho_{\rm DM}}{0.4 \, \text{GeV}/\text{cm}^3}\Big)^{1/2}$.

There exist many different types of experiments exploiting this electromagnetic mixing, including: cavities~\cite{Sikivie:1983ip,Rybka:2014xca,Woohyun:2016,Goryachev:2017wpw,Alesini:2017ifp,Melcon:2018dba,Melcon:2020xvj}, dielectric disks~\cite{TheMADMAXWorkingGroup:2016hpc,Baryakhtar:2018doz}, dish antennae~\cite{Horns:2012jf,Jaeckel:2013sqa,Suzuki:2015sza,Experiment:2017icw,BRASS}, plasmas~\cite{Lawson:2019brd}, LC circuits~\cite{Sikivie:2013laa,Chaudhuri:2014dla,Kahn:2016aff,Silva-Feaver:2016qhh,Crisosto:2018div}, and electric-field radios~\cite{Godfrey:2021tvs}. 

Cavities use the resonant enhancement of a cavity mode to increase the probability of DM converting to photons over a narrow frequency range, given by the so-called quality factor, that is the width of the resonant mode of the cavity. Contrary to cavities, dish antennae allow instead a broadband DM search, relying on the nonresonant breaking of translation invariance~\cite{Horns:2012jf,Jaeckel:2013sqa}. When the DPDM passes through the dish, the electrons in the antenna oscillate under the influence of the small dark photon electric field, and emit a (mostly) ordinary electromagnetic wave perpendicular to the surface of the antenna. Then, dielectric haloscopes rely on the same principles of a dish antenna, but are equipped with many semi-transparent dielectric layers, arranged to create constructive interference of the emitted waves, enhancing sensitivity at the expense of bandwidth. A distinct class of experiments, known as plasma haloscopes~\cite{Lawson:2019brd}, allow for dark photons converting to photons by matching the plasma frequency of specific material to the DM mass. Of course this could work for many plasmas, but a specific proposal, which allows for tuneable plasmas at the GHz regime, is to use thin aligned wire metamaterials~\cite{Lawson:2019brd}. Finally, LC circuits (lumped element circuits) try to measure a $B$-field, rather than the electric field. The $B$-field can be generated directly from the DM~\cite{Sikivie:2013laa}, or also indirectly via the $E$-field. This in fact can generate a displacement current $\textbf{J}_{X}$, which in turn leads to a magnetic field~\cite{Chaudhuri:2014dla,Kahn:2016aff,Silva-Feaver:2016qhh,Crisosto:2018div,Ouellet:2018beu}
\begin{subequations}
\begin{align}
    \nabla \times \textbf{B}_{A'} = \textbf{J}_{A'}  \, , \\
    \nabla \times \textbf{E}_{A'} = -\frac{\partial \textbf{B}_{A'}}{\partial t} \,,
\end{align}
    \end{subequations}
which will be read out by an inductive loop. The electromagnetic dual of a lumped element circuits experiment is to have a shielded room much larger than the Compton wavelength, and simply place an antenna to read out the electric field induced by DPDM~\cite{Godfrey:2021tvs}. Interestingly, dish antennas can also be used to look for the photons originating from the DPDM resonant conversion in close-by ionized plasmas, such as the Earth ionosphere~\cite{Beadle:2024jlr} or the solar corona~\cite{An:2020jmf, An:2023wij}.

\subsubsection{Dark matter energy injection at late times}
\vspace{0.1cm}

In this review, we have already encountered several times the concept of resonant conversion between photons and dark photons, which happens when, in a dilute QED plasma, the dark photon mass matches the plasma frequency, $\omega_{\rm pl} = m_{A'}$. We have seen, among other things, that such a resonance can lead to the copious production of dark photons in stellar interiors, $\text{plasma} \rightarrow A'$, as well as characteristic spectral distortions due to CMB photons disappearance, $\gamma_{\rm CMB} \rightarrow A'$. If dark photons are DM, then one can have the inverse process, with DPDM efficiently converting into SM photons $A'_{\rm DM} \rightarrow \gamma$ in plasma environments when a resonance is met. For example, resonant conversion of DPDM after recombination can produce excessive heating of the IGM, which is capable of prematurely reionizing hydrogen and helium. This heating can then leave a distinct imprint on both the Ly-$\alpha$ forest and the integrated optical depth of the CMB~\cite{McDermott:2019lch, Caputo:2020bdy, Bolton:2022hpt, Trost:2024ciu}. Such probes are particularly important for very small dark photon masses, $m_{A'} \lesssim 10^{-8}$~eV, the reason being twofold. First, it is for such small masses that resonance conversion happen at redshifts $z < 10^3$; second, the photons produced in the conversion are very soft, $\omega_\gamma \sim m_{A'}$, and can thus be absorbed very quickly by free-free absorption in the ionized IGM. The mean free path, $\lambda_{\rm ff}$, is given approximately by~\cite{Bolton:2022hpt}
\begin{equation}
\lambda_{\rm ff} \simeq \frac{14~\mathrm{kpc}}{(1 + z)^6} \, \Delta_b^{-2} 
\left( \frac{T}{10^4~\mathrm{K}} \right)^{3/2}
\left( \frac{m_{A'}}{10^{-13}~\mathrm{eV}} \right)^2\ ,
\end{equation}
where $T$ is the IGM temperature, while $\Delta_b \equiv \rho_b / \langle \rho_b \rangle$ is the local overdensity of baryons at which $A'$ conversion occurs. The mean free path $\lambda_{\rm ff}$ is much smaller than typical cosmological scales, thus one can assume that all the energy in the conversion is transformed into heating the medium. Now, the energy per baryon stored in the dark sector is quite large, $\rho_{\rm CDM}/n_b \sim \Omega_{\rm CDM}/\Omega_b \times m_p \sim 5 \times 10^9~\mathrm{eV}$, which implies a sizable energy injection \textit{per unit baryon} from DPDM resonant conversion~\cite{Bolton:2022hpt, Trost:2024ciu}
\begin{equation}
E_{A' \to \gamma} \sim \mathrm{eV} 
\left( \frac{\varepsilon}{10^{-15}} \right)^2 
\left( \frac{3}{1 + z_{\rm res}} \right)^{3/2} 
\left( \frac{\mdp}{10^{-13} \text{eV}} \right)\ ,
\end{equation}
where $z_{\rm res}$ is the redshift at which the resonance conversion happens, and where we assume the number density of baryons to evolve everywhere only through adiabatic expansion.

The same phenomenon of DPDM conversion into soft, quickly-absorbed photons can also happen in our local universe. In fact, one can use measurements from the gas cooling rate in our Milky Way~\cite{Dubovsky:2015cca} or ultra-faint dwarf galaxies, such Leo T~\cite{Wadekar:2019mpc} (there are no direct measurement for the heating rate, which is a bit unfortunate. One could then conservatively impose the heating rate due to dark matter to be smaller than the observed cooling rate). We notice, however, that these work rely on \textit{non-resonant} conversion, and the associated limits are weaker than the ones derived from cosmology. 

To illustrate that dark photons are very much still an active area of research, we point out that~\cite{Hook:2025pbn} recently raised doubts about the validity of dark-photon energy-injection bounds that rely on resonant conversion.  The central claim of that work is that nonlinear plasma effects become important and invalidate these constraints, down to kinetic mixing parameter values of order $10^{-14}$.  While the idea is very intriguing, we are not sure if the last word has been spoken about this, and hence we show the previous bounds for comparison as dashed lines in Fig.~\ref{fig:BoundsBelowMeV}.

\subsection{Dark photon as mediator to dark matter}\label{subsec:DMwithultralightDPmediator}

The dark photon can mediate interactions between the dark matter and the SM particles.  For very small dark photon masses, the phenomenology of such dark matter is very similar to those of millicharged particles, so that all the bounds and searches for millicharged particles are applicable.  
Much of the discussion in \S\ref{subsec:DPasmediatortoDM} also applies here, but there are some important difference, which we will briefly mention. 

If $\mdp$ is very small, the direct-detection scattering rate with, for example, electrons will be proportional to 
\begin{equation}
    \sigma_{\rm DD} \propto \frac{\alpha\alpha_D\epsilon^2\mu_{\chi,e}^2}{q^4}\,,
\end{equation}
where $q$ is the momentum transfer between the dark matter to the electron, and $\mu_{\chi,e}$ is the reduced mass of the dark matter and electron.  For low-threshold dark matter detectors, $q$ is very small and can greatly enhance the scattering rate~\cite{Essig:2011nj,Essig:2015cda}, which can lead to direct-detection experiments being sensitive even to dark matter that obtains its relic abundance through freeze-in (see discussion in \S\ref{subsubsec:DMDD-heavymediator}).  
At the same time, accelerator-based probes, which produce the dark matter with high energy, are less sensitive to such particles.  The region being probed by accelerator experiments is for relatively large values of $\epsilon$, which are so large that the scattering rate of dark matter with ordinary matter is too large to allow the dark matter to make it to terrestrial direct-detection experiments, as it will get stopped in the Earth's crust or atmosphere~\cite{Emken:2019tni}.  However, this parameter region with large $\epsilon$ can instead be probed through cosmology, as dark matter-baryon scattering in the early Universe would impact, e.g., the CMB~\cite{Boddy:2018wzy,Buen-Abad:2021mvc}, so only small fractions are allowed.  These can be probed with satellite-borne detectors~\cite{Alpine:2024kej,Du:2024afd}.  

If the dark photon mass is very small, then (besides allowing for dark matter self-interactions), it is also possible to have dissipative interactions among the dark matter particles.  Just like ordinary matter can dissipate energy whether it scatters by emitting photons, such dark matter can have interesting consequences for the formation of galactic structure, as it can lead, e.g., to the formation of exotic compact objects~\cite{Chang:2018bgx,Bramante:2024pyc}.

\section{Conclusions}
\label{sec:conclusions}
Dark photons have received a large amount of attention, especially over the past two decades. They are a relatively simple yet compelling extension to the Standard Model of particle physics, and they can impact a wide range of phenomena, including when they provide a connection between ordinary and dark matter.  This includes searches at colliders, fixed-target experiments, electron and proton beam-dump experiments, and direct-detection experiments, and they can impact precision measurements, as well as astronomical and cosmological observations.  Their theoretical importance and the fact that they have a rich phenomenology has contributed to their importance and popularity.  Although it seems that most of the phenomena have now been mapped out, they keep surprising us with additional discoveries being made.  In addition, they are often just one component, or take the place of a stand-in, of more complicated dark sectors. 

\begin{ack}[Acknowledgments]%
We thank  Maksym Ovchynnikov for providing the constraints from Ref.~\cite{Kyselov:2024dmi}. RE acknowledges support from DOE Grant DE-SC0025309, Simons Investigator in Physics Awards~623940 and MPS-SIP-00010469, Heising-Simons Foundation Grant No.~79921, and Binational Science Foundation Grant No.~2020220. AC is supported by an ERC STG grant (``AstroDarkLS'', grant No. 101117510). AC also acknowledges the Weizmann Institute of Science for hospitality at different stages of this project and the support from the Benoziyo Endowment Fund for the Advancement of Science.
\end{ack}

%%%%%%%%%%%%%%%%%%%%%%%%%%%%%%%%%%%%%%%%%%%%
%% Optional: A list of references to other relevant works/articles/websites which are not cited in the text but that would further enhance a readers understanding of this topic
%\seealso{article title article title}

%%%%%%%%%%%%%%%%%%%%%%%%%%%%%%%%%%%%%%%%%
%% Mandatory: Bibliography using bibtex 
\bibliographystyle{Numbered-Style} %% for Numbered Reference Style
\bibliography{reference}

\end{document}